\documentclass[letter,12pt]{article}
\usepackage{latexsym}
\usepackage{amsmath,amssymb,bm,ascmac,color,calc}
\usepackage{slashed}
\usepackage[mathscr]{eucal}
\usepackage{graphicx}
\usepackage{cite}
\usepackage{lineno}
\usepackage{jheppub}
\usepackage{multirow}
\usepackage{subfigure}
\usepackage[normalem]{ulem}

\usepackage{color}

\usepackage{makecell}

\allowdisplaybreaks

\newcommand{\ignore}[1]{}

\newcommand{\GeV}{{\rm\ GeV}}

\newcommand{\TeV}{{\rm\ TeV}}

\newcommand{\frules}{{\sc Feyn\-Rules}}

\newcommand{\onehalf}{\frac{1}{2}}
\newcommand{\threehalf}{\frac{3}{2}}

\def\beq{\begin{equation}}
\def\eeq{\end{equation}}
\newcommand{\ba}{\begin{array}}
\newcommand{\ea}{\end{array}}
\newcommand{\bea}{\begin{eqnarray}}
\newcommand{\eea}{\end{eqnarray} }
\newcommand{\bal}{\begin{align}}
\newcommand{\eal}{\end{align}}
\def\bi{\begin{itemize}}
\def\ei{\end{itemize}}
\def\ben{\begin{enumerate}}
\def\een{\end{enumerate}}
\def\beq{\begin{equation}}
\def\eeq{\end{equation}}
\def\bc{\begin{center}}
\def\ec{\end{center}}
\def\bt{\begin{table}}
\def\et{\end{table}}
\def\btb{\begin{tabular}}
\def\etb{\end{tabular}}

\title{Shedding Light on Top Partner at the LHC}

\author[a,b]{Haider Alhazmi,}
\author[a]{Jeong Han Kim,}
\author[a]{Kyoungchul Kong,}
\author[a]{Ian M. Lewis}

\affiliation[a]{Department of Physics and Astronomy, University of Kansas, Lawrence, KS 66045, USA}
\affiliation[b]{Department of Physics, Jazan University, Jazan 45142, Saudi Arabia}

\emailAdd{haider@ku.edu}
\emailAdd{kckong@ku.edu}
\emailAdd{jeonghan.kim@ku.edu} 
\emailAdd{ian.lewis@ku.edu} 

\abstract{

We investigate the sensitivity of the 14 TeV LHC to pair-produced top partners ($T$) decaying into the Standard Model top quark ($t$) plus either a gluon ($g$) or a photon ($\gamma$). The decays $T\rightarrow tg$ and $T\rightarrow t\gamma$ can be dominant when the mixing between the top partner and top quark are negligible.  In this case, the conventional decays $T\rightarrow bW$, $T\rightarrow tZ$, and $T\rightarrow th$ are highly suppressed and can be neglected.   We take a model-independent approach using effective operators for the $T$-$t$-$g$ and $T$-$t$-$\gamma$ interactions, considering both spin-$\frac{1}{2}$ and spin-$\frac{3}{2}$ top partners. 
 We perform a semi-realistic simulation with boosted top quark tagging and an appropriate implementation of a jet-faking-photon rate. 
Despite a simple dimensional analysis indicating that the branching ratios ${\rm BR}(T\rightarrow t\gamma)\ll {\rm BR}(T\rightarrow tg)$ due to the electric-magnetic coupling being much smaller than the strong force coupling,  
our study shows that the LHC sensitivity to $T\bar{T}\rightarrow t\overline{t}\gamma g$ is more significant than the sensitivity to $T\overline{T}\rightarrow t\overline{t}gg$.  This is due to much smaller backgrounds attributed to the isolated high-$p_T$ photon.  We find that with these decay channels and 3 ab$^{-1}$ of data, the LHC is sensitive to top partner masses $m_T\lesssim 1.4-1.8$~TeV for spin-$\frac{1}{2}$ and spin-$\frac{3}{2}$ top partners, respectively.}

\begin{document}

\maketitle

\section{Introduction} \label{sec:intro}
The Large Hadron Collider (LHC) is beginning to explore the physics of the TeV scale in earnest.  After the discovery of a Higgs boson, the next goal is to find new physics beyond the Standard Model (BSM).  There have been several different approaches in searches for new physics at the LHC.  One avenue that has not received as much attention is that the first discovery of BSM physics could arise from an effective field theory (EFT) that includes several new particles in addition to the SM~\cite{Craig:2016rqv}.  The number of theories of this type is, however, limited because new particles are identified by only a few quantum numbers (such as spin and gauge charge) which take only a small number of discrete values.

In this paper, we study vector-like fermionic top partners ($T$), whose left and right components have the same gauge quantum numbers as the top quark ($t$).  These particles appear in many BSM models as an attempt to cancel the quadratic divergences of the top quark loop contributions to the Higgs mass and stabilize the electroweak (EW) scale.  Such examples are Little Higgs models~\cite{ArkaniHamed:2002qy,ArkaniHamed:2002pa,Low:2002ws,Chang:2003un,Csaki:2003si,Perelstein:2003wd,Chen:2003fm,Berger:2012ec}, models with extra dimensions, composite Higgs models~\cite{Agashe:2004rs,Agashe:2005dk,Agashe:2006at,Contino:2006qr,Giudice:2007fh,Azatov:2011qy,Serra:2015xfa,Norero:2018dfg,Yepes:2017pjr,Sirunyan:2017tfc,Yepes:2018dlw}, etc.  To cure the ultraviolet sensitivity of the Higgs mass, top partners are postulated to be relatively light, with masses around the TeV range.  Many of these models involve a symmetry larger than the SM gauge symmetry, which implies that the new sector could be rich. However, given the current constraints from direct searches~\cite{Kim:2018mks,Sirunyan:2017yta,ATLAS-CONF-2018-032}, it is likely that we will have direct experimental access to only a subset of new particles.

In this paper, we examine non-standard decays of the top-partners that have often been neglected in LHC searches.  Typically, top partners are searched for in the three conventional decays $T\rightarrow bW$, $T\rightarrow tZ$, and $T\rightarrow th$~\cite{DeSimone:2012fs,Sirunyan:2017pks,Sirunyan:2017ynj,Sirunyan:2018omb,Aaboud:2017zfn,Aaboud:2017qpr,Aaboud:2018saj,Aaboud:2018xuw,ATLAS-CONF-2018-032}.  We focus on the top partner decays $T\rightarrow tg$ and $T\rightarrow t\gamma$.  The interactions $T-t-g$ and $T-t-\gamma$ does not appear at tree level due to gauge invariance, and therefore $T\rightarrow tg$ and $T\rightarrow t\gamma$ are typically suppressed relative to the conventional decays.  However, $T\rightarrow tg$ and $T\rightarrow t\gamma$ can be dominant when the mixing between the top partner and top quark is minimal~\cite{Kim:2018mks}.  We take a model-independent approach using effective operators between the top partner, top quark, and gauge bosons and consider both spin-$\frac{1}{2}$ and spin-$\frac{3}{2}$ top partners.  Searches for $T\rightarrow t\gamma$ have not been performed.  Additionally, while there have been searches for pair produced top partners decaying as $T\rightarrow tg$~\cite{Sirunyan:2017yta}, we update those analyses using boosted techniques and top-tagging of fat jets.   As we will show, although the $T\rightarrow t\gamma$ branching ratio is generically smaller than $T\rightarrow tg$ due to the gauge couplings, the LHC is more sensitive to  the signal $T\overline{T}\rightarrow t\overline{t}\gamma g$  than when both top partner decay into a top quark plus gluon.  This is due to the smaller backgrounds associated with requiring a hard isolated photon.

The rest of the paper is organized as follows.  We introduce two benchmark scenarios (spin-$\frac{1}{2}$ and spin-$\frac{3}{2}$) in section \ref{sec:model} and discuss production and decays of the top-partner in section \ref{sec:proddec}. Details of our analysis is presented in section \ref{sec:simulation}. Section \ref{sec:summary} is reserved for the summary and conclusions.  Further details of our collider analysis are given in the appendices.


\section{Theoretical Models} \label{sec:model}
We will denote the spin-$\frac{1}{2}$ top partner as $T_{\onehalf}$ and the spin-$\frac{3}{2}$ top partner as $T_{\threehalf}$.  The notation $T$ will be used for generic fermionic top partners.
 
\subsection{Spin-$\frac{1}{2}$ Top Partner}

We consider a model where the SM is extended with a spin-$\frac{1}{2}$ vector-like top partner ($T_{1/2}$) with an hypercharge of 2/3, which is a singlet under $SU(2)_L$ and a triplet under $SU(3)_C$. Its hypercharge, color representation, and spin determine its couplings to photons and gluons via its kinetic term
\begin{eqnarray}
\mathcal{L}_{kin} =  \overline{T}_{\onehalf}   \left( i \slashed{\partial} - g_1 Y_{t_R} \slashed{B} - g_3 T^A \slashed{G}^A\right) T_{\onehalf} ,
\end{eqnarray}
where $Y_{t_R} = Q_t = \frac{2}{3}$ is the $U(1)_Y$ hypercharge of the right-handed top quark, and $T^A$'s are the fundamental generators of $SU(3)_C$.

Since the top partner and the SM top quark have the same unbroken quantum numbers they can mix, giving rise to additional couplings with EW gauge bosons and the Higgs. As a result, the most commonly studied $T_{\onehalf}$ decay modes are $t Z$, $t h$ and $Wb$ where the exact decay rates are determined by the mixing angle and model parameters.  While in principle there are two mixing angles, one each for the left- and right-components of the top quark and top partner, there are only three mass terms in the Lagrangian~\cite{Kim:2018mks}
\begin{eqnarray}
\mathcal{L}_{\rm mass}=-y_t \overline{Q_L}\Phi\,t_R-\lambda_t\overline{Q_L}\widetilde{\Phi}\,T_{\onehalf,R}-m_2\,\overline{T}_{\onehalf,L}T_{\onehalf,R}+{\rm h.c.},
\end{eqnarray}
where $Q_L$ is the third generation $SU(2)_L$ quark doublet, $\Phi$ is the Higgs doublet, and $\widetilde{\Phi}=\varepsilon \Phi^*$.   Hence, there are only three free parameters: the top quark mass $m_t$, the top partner mass $m_T$, and one mixing angle which we choose to be the left-handed mixing angle $\theta_L$. Relationships between the masses, mixing angles, and Lagrangian parameters can be found in Ref.~\cite{Kim:2018mks}.

The mixing angle $\theta_L$ is highly constrained by EW precision measurements~\cite{He:2001tp,Chen:2017hak,Chen:2014xwa,Dawson:2012di, Aguilar-Saavedra:2013qpa}. The oblique parameters constrain $|\sin\theta_L|\lesssim 0.16$ for the top partner mass around 1 TeV and $|\sin\theta_L|\lesssim 0.11$ for $m_T\gtrsim 2$ TeV~\cite{Chen:2017hak,Dawson:2012di, Aguilar-Saavedra:2013qpa}.  Measurement of the CKM matrix element $|V_{tb}|=1.019\pm0.025$~\cite{PDG2018} can also constrain the mixing parameter to be $|\sin\theta_L| <  0.11$ independent of $m_T$, which is comparable to the EW precision measurements. The collider bounds turn out to be less constraining~\cite{ATLAS-CONF-2016-072}. Hence, it is essential to scrutinize the parameter space where the mixing angle goes to zero and the conventional tree-level decays $T_{\onehalf} \rightarrow t Z$, $T_{\onehalf} \rightarrow t h$ and $T_{\onehalf} \rightarrow bW$ vanish.

In a model with a SM gauge singlet scalar ($S$) in addition to the top partner, it is possible that the scalar $S$ can induce new loop level decays $T_{\onehalf} \rightarrow t g$, $T_{\onehalf} \rightarrow t \gamma$, and $T_{\onehalf} \rightarrow t Z$~\cite{Kim:2018mks,Freitas:2017afm}. All these modes survive even in the zero-mixing limit while other tree-level decay modes are closed. In this case, the branching ratios are mostly determined by the gauge couplings and weak mixing angle. Among these, the decay $T_{\onehalf} \rightarrow t g$ is expected to be dominant due to the strong coupling, while the other decays $T_{\onehalf} \rightarrow t \gamma$ and $T_{\onehalf} \rightarrow t Z$ would be suppressed by the weak couplings. 
Let us consider the following dipole operators in the limit where the scalar $S$ is integrated out,  
\begin{eqnarray}
\label{eq:EFT0} \mathcal{L}_{EFT} &=& \frac{c_{3} g_3 }{\Lambda} \overline{T}_{\onehalf}\sigma^{\mu\nu}T^A t_R G^{A}_{\mu\nu} + 
\frac{c_{1} g_1 }{\Lambda} \overline{T}_{\onehalf}\sigma^{\mu\nu} Y_{t_R} t_R B_{\mu\nu} + {\rm h.c.} \,\label{eq:EFTSU2} . ~~~~
\end{eqnarray}
where $\Lambda$ is a heavy new physics scale, $\sigma^{\mu\nu}=\frac{i}{2}[\gamma^\mu,\gamma^\nu]$, $G^{A}_{\mu\nu}$ is the gluon field strength tensor, and $B_{\mu\nu}$ is the hypercharge field strength tensor. The couplings $g_3$ and $g_1$ are the strong and hypercharge couplings, respectively.  After EW symmetry breaking (EWSB), Eq. (\ref{eq:EFT0}) can be effectively parameterized by the short-distance interactions between the top partner and top mediated by a gluon, photon or $Z$ as in Eq.(\ref{eq:EFT}). 
\begin{eqnarray}
\mathcal{L}_{EFT}= c_{g} \overline{T}_{\onehalf}\sigma^{\mu\nu}T^A t_R G^{A}_{\mu\nu} + c_{\gamma} \overline{T}_{\onehalf}\sigma^{\mu\nu}t_R F_{\mu\nu} + c_{Z} \overline{T}_{\onehalf}\sigma^{\mu\nu}t_R Z_{\mu\nu}  + {\rm h.c.} \, ,\label{eq:EFT}
\end{eqnarray}
where 
\begin{eqnarray}
c_g = \frac{c_{3} g_3 }{\Lambda},\quad c_\gamma = \frac{c_1}{\Lambda} Q_{t} e \cos\theta_W,\quad c_Z = \frac{c_1}{\Lambda}  Q_{t} e\,\tan\theta_W,\label{eq:coeff}
\end{eqnarray}
 $\theta_W$ is the weak mixing angle, and $Q_t=\frac{2}{3}$ is the top quark's electric charge.
The field strength tensors for gluon, photon and $Z$ are defined as
\begin{eqnarray}
G_{\mu\nu}^A&=& \partial_\mu G^A_\nu-\partial_\nu G^A_\mu-g_3 f^{ABC}G^B_\mu G^C_\nu \, ,\label{eq:fieldstrength1} \\
F_{\mu\nu}&=&\partial_\mu F_\nu-\partial_\nu F_\mu \, , \\
Z_{\mu\nu} &=&\partial_\mu Z_\nu-\partial_\nu Z_\mu \, ,\label{eq:fieldstrength3}
\end{eqnarray}
where $f^{ABC}$ is the $SU(3)_C$ structure constant. This gives rise to similar decay patterns of excited quarks discussed in Refs.~\cite{DeRujula:1983ak,Kuhn:1984rj,Baur:1987ga,Baur:1989kv,Han:2010rf}.

\subsection{Spin-$\frac{3}{2}$ Top Partner}

Spin-$\frac{3}{2}$ fermions are described by the Rarita-Schwinger Lagrangian, which is a generalized version of the Dirac equation with a Lorentz index on a Dirac spinor \cite{Rarita:1941mf}. In this section, we make brief remarks on interactions that are relevant  for the rest of this paper.  Please refer to to Refs. \cite{Rarita:1941mf,Hassanain:2009at,Dicus:2012uh,Moussallam:1989nm,Stirling:2011ya,Christensen:2013aua} for more details on the physics of spin-$\frac{3}{2}$ top partners. 

The interaction of spin-$\frac{3}{2}$ top partner ($T_{\threehalf}^\alpha$) and the SM gluon ($G^A_\mu $) is given by
\begin{equation}
{\cal L} \ni g_3 \overline{T}^\alpha_{\threehalf} \left ( \frac{3z^2+2z+1}{2} \gamma_\alpha \gamma_\mu \gamma_\beta + z g_{\alpha\mu}\gamma_\beta + z \gamma_\alpha g_{\mu\beta} + g_{\beta\alpha}\gamma^\mu  \right ) T^A T_{\threehalf}^\beta G^{A,\mu} \, ,\label{eq:T32Kin}
\end{equation}
where the $z$ is an unphysical, arbitrary parameter ($z \neq-\frac{1}{2}$). In principle, all physical quantities should be independent of this parameter \cite{Christensen:2013aua,Rarita:1941mf,Hassanain:2009at,Dicus:2012uh,Moussallam:1989nm,Stirling:2011ya}.
The spin-$\frac{3}{2}$ particle cannot mix with the top quark, and its decays are described by an EFT.  For an $SU(2)_L$ singlet $T_{\threehalf}$, the effective Lagrangian describing the interaction between $T_{\threehalf}$, the SM top quark, and gauge bosons is
\begin{eqnarray}
\mathcal{L}_{EFT}&=&i\,\frac{g_3c_3}{\Lambda}\, \overline{T}_{\threehalf}^\mu\,\left(g_{\mu\alpha}+z\,\gamma_\mu\gamma_\alpha\right)\,\gamma_\beta\,T^A\,t_R G^{A,\alpha\beta}+i\,\frac{g_1c_1}{\Lambda}\, \overline{T}_{\threehalf}^\mu\,\left(g_{\mu\alpha}+z\,\gamma_\mu\gamma_\alpha\right)\,\gamma_\beta\,t_R B^{\alpha\beta}\nonumber\\
&&+\rm{h.c.}~\label{eq:32EFT1}
\end{eqnarray}
In principle, the Wilson coefficients $c_{1,3}$ and scale of new physics $\Lambda$ are different than those in Eq.~(\ref{eq:EFTSU2}), but for simplicity we use the same notation.  Similarly, the parameter $z$ in Eq.~(\ref{eq:32EFT1}) can be different for each gauge boson before EWSB and different than the $z$ in Eq.~(\ref{eq:T32Kin}), but since they are unphysical they can be set equal without loss of generality.  After EWSB, the Lagrangian is
\begin{eqnarray}
\mathcal{L}_{EFT}&=&i\,\frac{c_g}{\Lambda}\, \overline{T}_{\threehalf}^\mu\,\left(g_{\mu\alpha}+z\,\gamma_\mu\gamma_\alpha\right)\,\gamma_\beta\,T^A\,t_R G^{A,\alpha\beta}+i\,\frac{c_\gamma}{\Lambda}\, \overline{T}_{\threehalf}^\mu\,\left(g_{\mu\alpha}+z\,\gamma_\mu\gamma_\alpha\right)\,\gamma_\beta\,t_R F^{\alpha\beta}\nonumber\\
&&+i\,\frac{c_Z}{\Lambda}\, \overline{T}_{\threehalf}^\mu\,\left(g_{\mu\alpha}+z\,\gamma_\mu\gamma_\alpha\right)\,\gamma_\beta\,t_R Z^{\alpha\beta}+{\rm h.c},\label{eq:32EFT}
\end{eqnarray}
where $c_\gamma$, $c_Z$, $c_g$ are given in Eq.~(\ref{eq:coeff}), and the field strength tensors are given in Eqs.~(\ref{eq:fieldstrength1}-\ref{eq:fieldstrength3}).


\section{Production and Decay} \label{sec:proddec}
\begin{figure}[tb]
\begin{center}
\includegraphics[width=0.31\textwidth, clip]{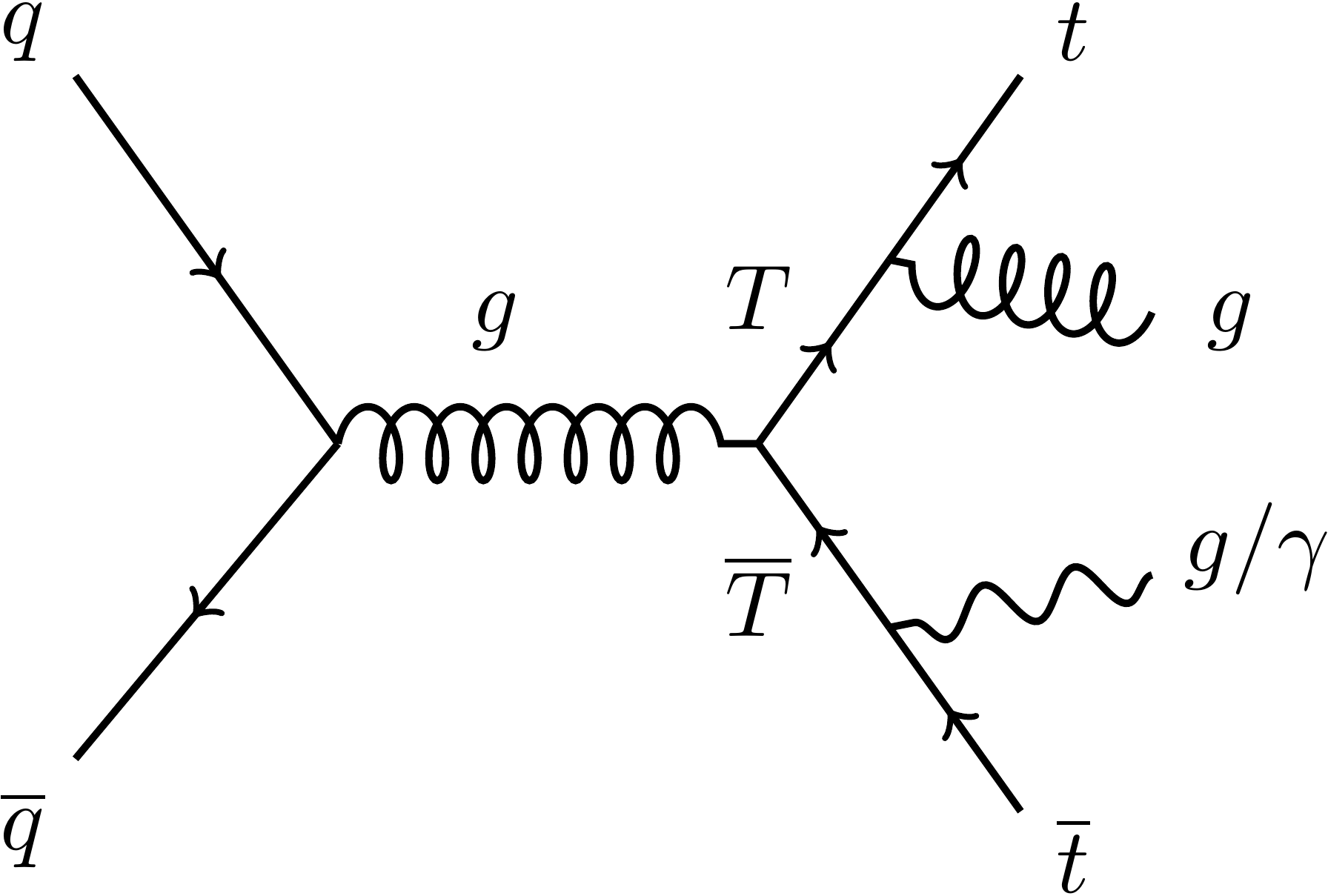} \hspace*{0.5cm}
\includegraphics[width=0.31\textwidth, clip]{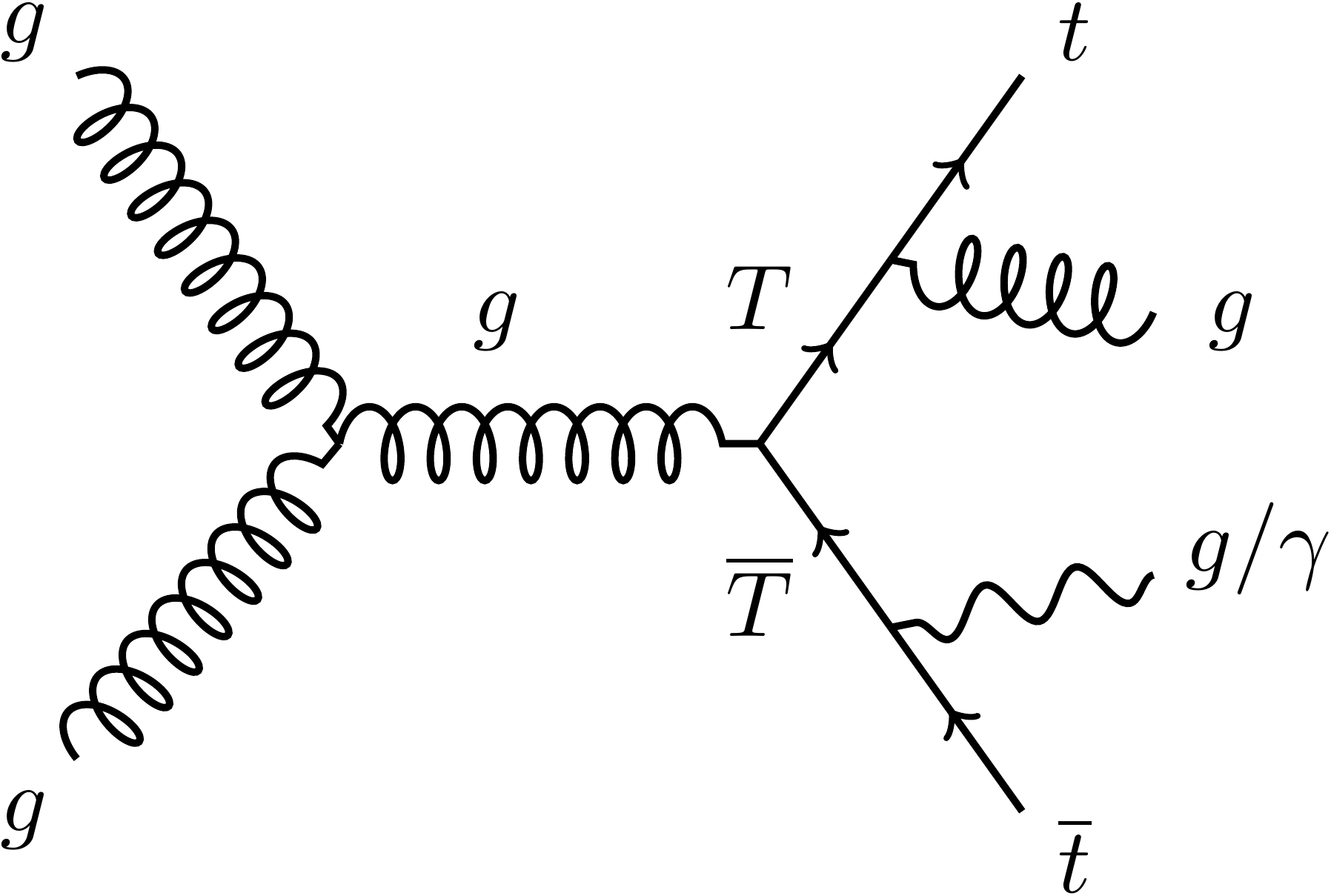} \hspace*{0.5cm}
\includegraphics[width=0.24\textwidth, clip]{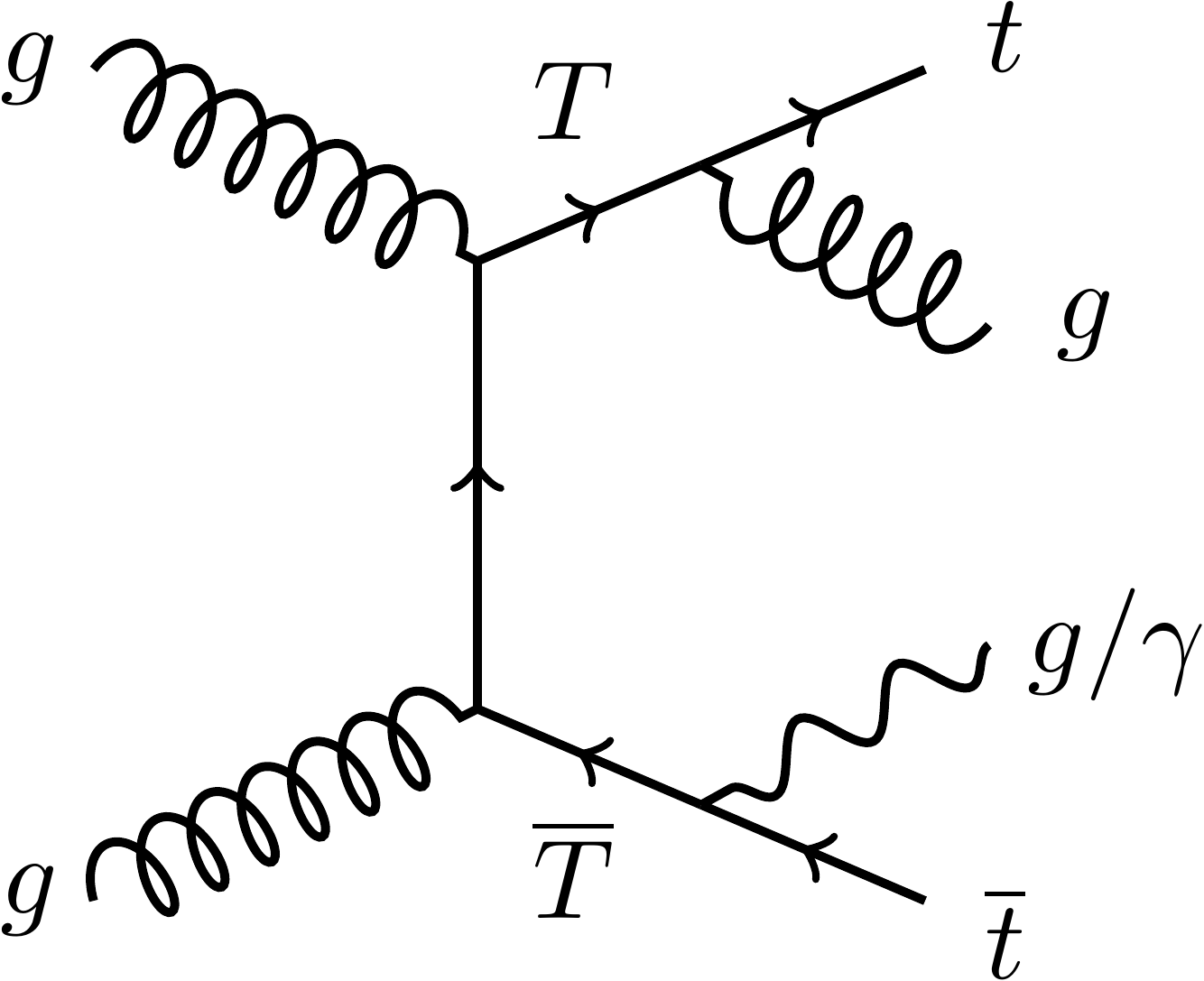}
\end{center}
\caption{Feynman diagrams for pair production of top partners at the LHC and their decays into $T \overline{T} \rightarrow t \overline{t} + g +  g  / \gamma$ final states.\label{fig:Pro}}
\end{figure}

\begin{figure}[tb]
\centering
\includegraphics[width=0.65\textwidth,clip]{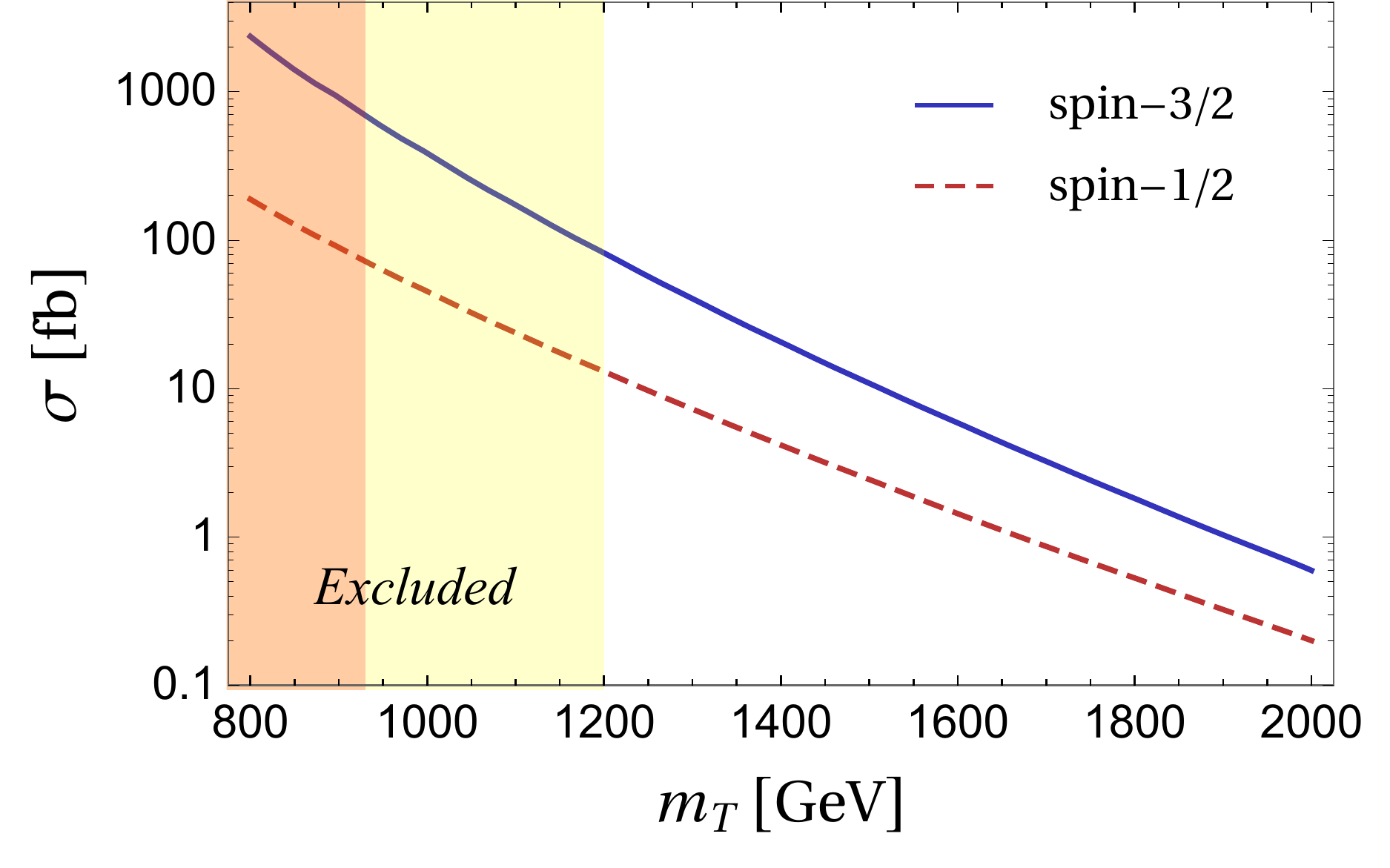}
\vspace*{-0.3cm}
\caption{\label{fig:xsection1}Pair production cross sections of spin-$\frac{1}{2}$ (red, dashed) and spin-$\frac{3}{2}$ (blue, solid) top partners at leading order accuracy as a function of the top partner mass ($m_T$) at the 14 TeV LHC. Yellow (red) shaded regions represents the spin-$\frac{3}{2}$ (spin-$\frac{1}{2}$) top partner masses excluded by 13 TeV data at 95\% C.L. with 35.9 fb$^{-1}$.
}
\end{figure}

We now discuss the production and decay of the top partner, $T$, introduced in Section~\ref{sec:model}. Interactions between color-triplet spin-$\frac{3}{2}$ and spin-$\frac{1}{2}$ particles and the SM gluon are fixed by $SU(3)_C$ gauge invariance. The Feynman diagrams for top partner pair production, $pp\rightarrow T\overline{T}$, are shown in Fig.~\ref{fig:Pro}.  Since the spin-$\frac{1}{2}$ top partner has the exact same color, spin, and electromagnetic quantum numbers of the top quark, pair production of $T_{\onehalf}$ is identical to that of SM top quark production with different masses.  Since the spin-$\threehalf$ top partner has different spin,  production of $T_{\threehalf}\overline{T}_{\threehalf}$ requires a careful calculation using the interactions in Eq.~(\ref{eq:T32Kin}).   In Fig.~\ref{fig:xsection1}, we show the pair production cross section at leading order (LO) as a function of the mass of $T$ for both spin-$\frac{3}{2}$ (blue, solid) and spin-$\frac{1}{2}$ (red, dashed).  The cross section was obtained using \texttt{MadGraph5\_aMC@NLO}~\cite{Alwall:2014hca} at the 14 TeV LHC with default parton distribution functions \texttt{NNPDF2.3QED}~\cite{Ball:2013hta}. For spin-$\frac{3}{2}$, we use the existing model file described in Ref. \cite{Christensen:2013aua} and have cross-checked these results using \texttt{CalcHEP} with our own implementation \cite{Belyaev:2012qa}. We also verified analytically that the pair production cross section of the spin-$\frac{3}{2}$ top partner agreed with the results in Ref. \cite{Dicus:2012uh}.

 A search for pair production of spin-$\threehalf$ vector-like quarks, each decaying exclusively to a top quark and a gluon, was recently performed by CMS~\cite{Sirunyan:2017yta} at the 13 TeV LHC with 35.9 fb$^{-1}$. Assuming ${\rm BR}(T\rightarrow tg) = 1$, a traditional analysis based on slim jets excluded masses below $ \sim 1.2$ TeV.   Recasting the CMS search~\cite{Sirunyan:2017yta} to a bound on the spin-$\onehalf$ top partner, a lower limit on the mass is found to be $m_T\gtrsim 930$ GeV.  The NNLO pair production cross section of $T_{\onehalf}$~\cite{Sirunyan:2017usq, Czakon:2011xx, Czakon:2013goa, Czakon:2012pz, Czakon:2012zr, Cacciari:2011hy} was used for the recast.  The lower limits on the masses of $T_{\threehalf}$ and $T_{\onehalf}$ are shown as the yellow and red shaded regions in Fig.~\ref{fig:xsection1}, respectively. 

While the $T\rightarrow tg$ decay is expected to be dominant due to the strong coupling in $g_3$ in Eqs.~(\ref{eq:EFT},\ref{eq:32EFT}), there is a non-negligible partial decay width to $t \gamma$.  For the case with $c_1= c_3 = 1$, we obtain the following partial decay widths
\begin{eqnarray}
\Gamma &=& \frac{{\cal C}}{16 \pi} \frac{{m_T}^3}{\Lambda^2} \left ( 1- \frac{{m_t}^2}{{m_T}^2} \right )^3 \, , \hspace*{3.3cm}{\rm for~ spin-}\frac{1}{2} \, ,\\
\Gamma &=& \frac{{\cal C}}{48 \pi} \frac{{m_T}^3}{\Lambda^2} \left ( 1- \frac{{m_t}^2}{{m_T}^2} \right )^3 \left ( 3 + \frac{{m_t}^2}{{m_T}^2} \right ) \, , \hspace*{1.cm}{\rm for~ spin-}\frac{3}{2} \, , 
\end{eqnarray}
 where the coefficient ${\cal C}$ is
 \begin{eqnarray}
 {\cal C}= \left \{
 \begin{array}{ll}
 (g_1 Y_{t_R} \cos\theta_W)^2 = (e Q_t)^2 				& ~~~{\rm for}~  \gamma \, t \\
 (g_1 Y_{t_R} \sin\theta_W)^2 = (e Q_t  \tan\theta_W)^2 \, , 	& ~~~{\rm for}~Z \, t    ~({\rm in~ the~} M_Z \ll m_T ~{\rm limit}) \, , \\
 g_3^2 C_2(R) = \frac{4}{3} g_3^2						&~~~ {\rm for}~ g  \, t
 \end{array}
 \right. 
 \end{eqnarray}
and $C_2(R)=\frac{4}{3}$ is the eigenvalue of the quadratic Casimir operator of the fundamental representation of $SU(3)_C$.
Therefore the partial widths are given by ratios of above coefficients:
\begin{eqnarray}
\Gamma( T  \to t\,\gamma)  :  \Gamma( T  \to t\,Z) :  \Gamma( T \to t\,g) &=& 
(e Q_t)^2  :   (e Q_t  \tan\theta_W)^2  :  g_3^2 C_2(R)
\end{eqnarray}
and the branching ratios are
\begin{eqnarray}
{\rm BR}(T\rightarrow t\gamma)=0.021,\quad {\rm BR}(T\rightarrow tZ)=0.0060,\quad {\rm BR}(T\rightarrow tg)=0.97,\label{eq:BR}
\end{eqnarray}
where $g_3$ is evaluated at two loops and the scale $1$~TeV.  These branching ratios are independent of the top partner spin.

Since the branching fraction of $T \to t Z$ is negligible, we focus on two other decay modes in this study. 
The independent parameters of the model are then
\begin{eqnarray}
c_{g}, c_{\gamma}, \,{\rm and}\,~ m_T.
\end{eqnarray}
%


\section{Searches for Top Partners at the LHC}\label{sec:simulation}

In this section, we perform a detailed collider analysis at the 14 TeV LHC.  We will take the spin-$\frac{1}{2}$ top partner as our benchmark model. 
Results for the spin-$\threehalf$ top partner will be discussed in section \ref{sec:results}. We consider the QCD pair production of a TeV scale top partner $T_{\onehalf}$ decaying into two final states
\bea
  p ~ p \rightarrow T_{\onehalf} ~ \overline{T}_{\onehalf} \rightarrow t ~ \overline{t} + g +  g  / \gamma \, , \label{eq:signal}
\eea 
where the tops are forced to decay semi-leptonically to avoid QCD multi-jet backgrounds. Additionally, since $m_T\gg m_t$ the top quarks are boosted.  Final states are, therefore, characterized by two boosted tops in association with two hard jets ($t \overline{t} g g$) or a hard jet and an isolated photon ($t \overline{t} g \gamma$). The $t \overline{t} g \gamma$ channel renders a relatively clean final state with small backgrounds, while the $t \overline{t} g g$ channel has a busy environment with a large irreducible $t \overline{t}$ background. 

The overall sensitivities of two channels depend on the branching ratios of $T_{\onehalf}$.  We will be interested in top partner masses of $m_T\sim 1-2$~TeV.  In this mass range, the branching ratios in Eq.~(\ref{eq:BR}) are insensitive to the running of $g_3(m_T)$.  Hence, for a benchmark point to determine the mass reach of the LHC, we will take $c_1=c_3=1$ where the top partner branching ratios are
\begin{eqnarray}
{\rm BR}(T\rightarrow t\gamma)\approx 0.03,\quad{\rm and}\quad {\rm BR}(T\rightarrow tg)\approx 0.97.\label{eq:benchmark}
\end{eqnarray}
As mentioned previously, the decay $T\rightarrow tZ$ has been neglected since its branching ratio is sub-percent level, as shown in Eq.~(\ref{eq:BR}).  For different branching ratios, our analysis can be simply rescaled as long as the total width of the top partner ($\Gamma_T$) is sufficiently small, $\Gamma_T/m_T\ll1$, and the narrow width approximation is valid. We will generalize the assumption on the branching ratio as a function of ${\rm BR} ( T \rightarrow t \gamma )$ in later discussions for a more comprehensive prediction. 

The models in section~\ref{sec:model} are implemented into the \frules~package~\cite{Alloul:2013bka}, which is in turn used to generate a UFO library~\cite{Degrande:2011ua} for \texttt{MadGraph5\_aMC@NLO}~\cite{Alwall:2014hca}. Both signal and background events are simulated by \texttt{MadGraph5\_aMC@NLO} at $\sqrt{S} = 14$ TeV using the default \texttt{NNPDF2.3QED}~\cite{Ball:2013hta} parton distribution functions. 
We use default dynamic renormalization and factorization scales. At particle-level, for both $t \overline{t} g g$ and $t \overline{t} g \gamma$ channels, we require all partons to pass the following cuts 
\begin{eqnarray}
p_T > 30~{\rm GeV},\quad{\rm and}\quad ~| \eta |< 5,\label{eq:basepartons}
\end{eqnarray}
 while leptons are required to have 
\begin{eqnarray}
p_T^\ell > 30~{\rm GeV}\quad{\rm and}\quad ~| \eta^\ell |< 2.5,\label{eq:baseleptons}
\end{eqnarray}
where $p_T$ are transverse momentum, $\eta$ is rapidity, and $\ell$ indicates leptons. To improve the statistics in the SM backgrounds, we demand 
\begin{eqnarray}
H_T > 700 \GeV,\label{eq:baseht}
\end{eqnarray} 
where $H_T$ denotes the scalar sum of the transverse momenta of all final state partons (excluding leptons and photons). On top of the generation-level cuts in Eqs.~(\ref{eq:basepartons}-\ref{eq:baseht}), an additional photon selection is required for the $t \overline{t} g \gamma$ channel with the photon passing the cuts 
\begin{eqnarray}
p_T^\gamma > 300~{\rm GeV}\quad{\rm and}\quad ~| \eta^\gamma |< 2.5. \label{eq:basephoton}
\end{eqnarray}
All the particle-level events are showered and hadronized by \texttt{PYTHIA6}~\cite{Sjostrand:2006za} and clustered by the \texttt{FastJet}~\cite{Cacciari:2011ma} implementation of the anti-$k_T$ algorithm~\cite{Cacciari:2008gp} with a fixed cone size of $r = 0.4 \;(1.0)$ for a slim (fat) jet.  We match the hadronized and showered event using the MLM method~\cite{MLM}.

We also include simplistic detector resolution effects based on the ATLAS detector performances~\cite{ATL-PHYS-PUB-2013-004, Aad:2014nim}, and smear momenta and energies of reconstructed jets, photons and leptons according to the value of their energies, as described in appendix \ref{app:DR}.


\subsection{$t \overline{t} g g$ Decay Channel}\label{sec:analysis1}
We now focus on the $t \overline{t} g g$ channel and describe our analysis strategy to estimate the LHC sensitivity to this channel. The previous CMS search~\cite{Sirunyan:2017yta} for $T$ in the $t \overline{t} g g$ channel utilized exactly one isolated lepton, $\slashed{E}_T$, at least six slim jets, and exactly two $b$-tagged jets, to take into account the busy environment. The main challenge of such a high jet-multiplicity environment is to resolve the combinatorial problem of determining which final state objects originated from which decaying particles and fully reconstruct the event.  A typical method is to employ a chi-square fit to reconstruct the masses of the $W$ bosons, top quarks and top partners using truth information from simulated signal samples. The success rate of accurately reconstructing all objects was found to be only $11\%$~\cite{Sirunyan:2017yta}.

When searching for top partners with masses at the TeV scale, the character of signal events changes.  The top quarks are boosted and their decay products highly collimated.  Additionally, the gluons from heavy $T$ decays are harder than those typically produced by QCD.  It is clear that the combinatorics becomes much simpler, but on the other hand, a traditional slim-jet-based analysis is no longer adequate.  Since the top quark decay products are highly collimated, they can be better clustered by fat jets with unique internal substructures. In this case, the jet-substructure analysis becomes more efficient compared to the conventional approach. 

As an extension to CMS study~\cite{Sirunyan:2017yta}, we present a new jet substructure analysis focusing on the boosted parameter space. We will demonstrate that the method can improve the resolution of the reconstructed $T$ mass.  This plays an important role in disentangling the signal from the irreducible $t \overline{t}$ background and enhancing the final signal sensitivity.

The $t \overline{t} g g$ channel has large SM backgrounds. 
The dominant background is semi-leptonic $t \overline{t}$ matched up to two additional jets. 
The single-top processes include $tW$ and $tq$, where $q$ is a light quark or a $b$-quark.  For the $tW$ background, one $W$ decays leptonically, one $W$ decays hadronically, and we match with up to three additional jets. The $tq$ process is matched with up to two additional jets and we only consider a top quark which decays leptonically. The sub-leading background includes $W$ matched with up to four additional jets where $W$ is decayed leptonically. 
The other insignificant backgrounds include $W W$ matched with up to three additional jets, where one $W$ decays leptonically and the other hadronically. The $W Z$ background is generated with up to three additional jets where the $W$ is forced to decay leptonically and the $Z$ hadronically.  

Table \ref{tab:TotalBackG} summarizes the background simulations, including detailed matching schemes, with the generation-level cuts in Eqs.~(\ref{eq:basepartons}-\ref{eq:baseht}).  To validate our background simulation, we reproduced the total number of background events in CMS~\cite{Sirunyan:2017yta} at a $\sqrt{S} = 13$ TeV using the same cut-based analysis. Our results for both the $\mu+{\rm jets}$ and $e+{\rm jets}$ were within $4\%$ agreement with the CMS simulations.  We have also confirmed good agreement with background estimation in the 8 TeV CMS analysis \cite{Chatrchyan:2013oba}.
\begin{table}[tb]
\begin{center}
\scalebox{1.0}{
\begin{tabular}{|c|c|c|c|c|}
\hline
           Abbreviations           & Backgrounds                       & Matching       & $\sigma \cdot {\rm BR~(fb)}$     \\ \hline	
  $t \overline{t} $                 &  $t \overline{t} + \rm{jets}$   &   4-flavor        &$  2.91 \times 10^{3} \; \rm fb $    \\ \hline
 \multirow{2}{*}{Single $t$} & $t W + \rm{jets}$                  &    5-flavor      &$  4.15 \times 10^{3} \; \rm fb $      \\
                                           & $tq+{\rm jets}$                      &    4-flavor      & 	$ 77.2 \; \rm fb$                             \\ \hline
  $W$                                  &$W + \rm{jets}$                     &     5-flavor     &   $ 4.96  \times 10^{3} \; \rm fb $     \\ \hline  
  \multirow{2}{*}{$VV $}      & $W W + \rm{jets}$                &    4-flavor      &  $ 111 \; \rm fb $                                  \\ 
                                           &$W Z  + \rm{jets}$                 &    4-flavor      &  $ 43.5  \; \rm fb$                             \\

\hline		
\end{tabular}}
\end{center}
\caption{The summary of the SM backgrounds relevant to the $t \overline{t} g g$ channel and their cross sections after generation level cuts in  Eqs.~(\ref{eq:basepartons}-\ref{eq:baseht}). Matching refers to either the 4-flavor or 5-flavor MLM matching~\cite{MLM}. The last column $\sigma \cdot \rm BR$ denotes the production cross section (in fb) times branching ratios including the top, $W$, and $Z$ decays.}
\label{tab:TotalBackG}
\end{table}

We now present detailed event selection cuts. Since the signal events contain one boosted leptonic top $t\rightarrow b\ell\nu$, our base-line selection cuts start from requiring a missing transverse energy of
\begin{eqnarray}
\slashed{E}_T > 50 \GeV\label{eq:ETmiss},
\end{eqnarray}
at least one $r = 0.4$ slim jet with 
\begin{eqnarray}
p_T^j > 30 \GeV\quad{\rm and}\quad|\eta^j| < 2.5,\label{eq:basethinjet}
\end{eqnarray}
and exactly one isolated lepton passing the cuts in Eq.~(\ref{eq:baseleptons}) and
\begin{eqnarray}
 p_T^\ell / p^{\Sigma}_T > 0.7,\label{eq:isolepton}
\end{eqnarray}
where $p^{\Sigma}_T$ is the scalar sum of transverse momenta of final state particles (including the lepton itself) within $\Delta R = 0.3$ isolation cone\footnote{The angular distance $\Delta R_{ij}$ is defined by
$\Delta R_{ij} = \sqrt{(\Delta\phi_{ij})^2+(\Delta \eta_{ij})^2}$, where $\Delta\phi_{ij} = \phi_i-\phi_j$ and $\Delta\eta_{ij}=\eta_i-\eta_j$ are the differences of the azimuthal angles and rapidities between particles $i$ and $j$ respectively.}.
At least one hard fat jet with
\begin{eqnarray}
p_T^j > 350 \GeV\quad{\rm and}\quad |\eta^j| < 2.5 , \label{eq:basefatjet}
\end{eqnarray}
is required to account for the boosted hadronic top candidate. The series of cuts described in Eqs.~(\ref{eq:ETmiss}-\ref{eq:basefatjet}) define our basic cuts for the $t \overline{t} g g$ channel.

Since the decay products of the boosted hadronic top are highly-collimated, we will identify the fat jet with a three-pronged substructure as the hadronic top candidate. This feature is distinguished from QCD jets, which typically have a two-pronged topology.  Therefore the SM backgrounds without a hard hadronic top can be substantially vetoed. We use the \texttt{TemplateTagger v.1.0}~\cite{Backovic:2012jk} implementation of the Template Overlap Method (TOM)~\cite{Almeida:2010pa,  Backovic:2013bga} to tag massive boosted objects\footnote{For comparisons with other popular taggers, see Ref.~\cite{Almeida:2015jua} and references therein.}. 
The TOM aims to match the energy distribution of a fat jet to three-pronged templates by scanning over the allowed phase space with all relevant kinematic constraints.  The likelihood of a fat jet originating from a parent particle $a$ with an $i$-pronged decay is encoded in an overlap score $Ov_i^a$.     Fat jets that are likely to have originated from the particle $a$ have an overlap score $Ov_i^a$ nearer one, while those that are unlikely to have originated from $a$ have $Ov_i^a$ closer to zero.  This method is not very susceptible to pileup contamination~\cite{Backovic:2013bga}. 

\begin{figure}[tb]
\centering
\includegraphics[width=0.49\textwidth,clip]{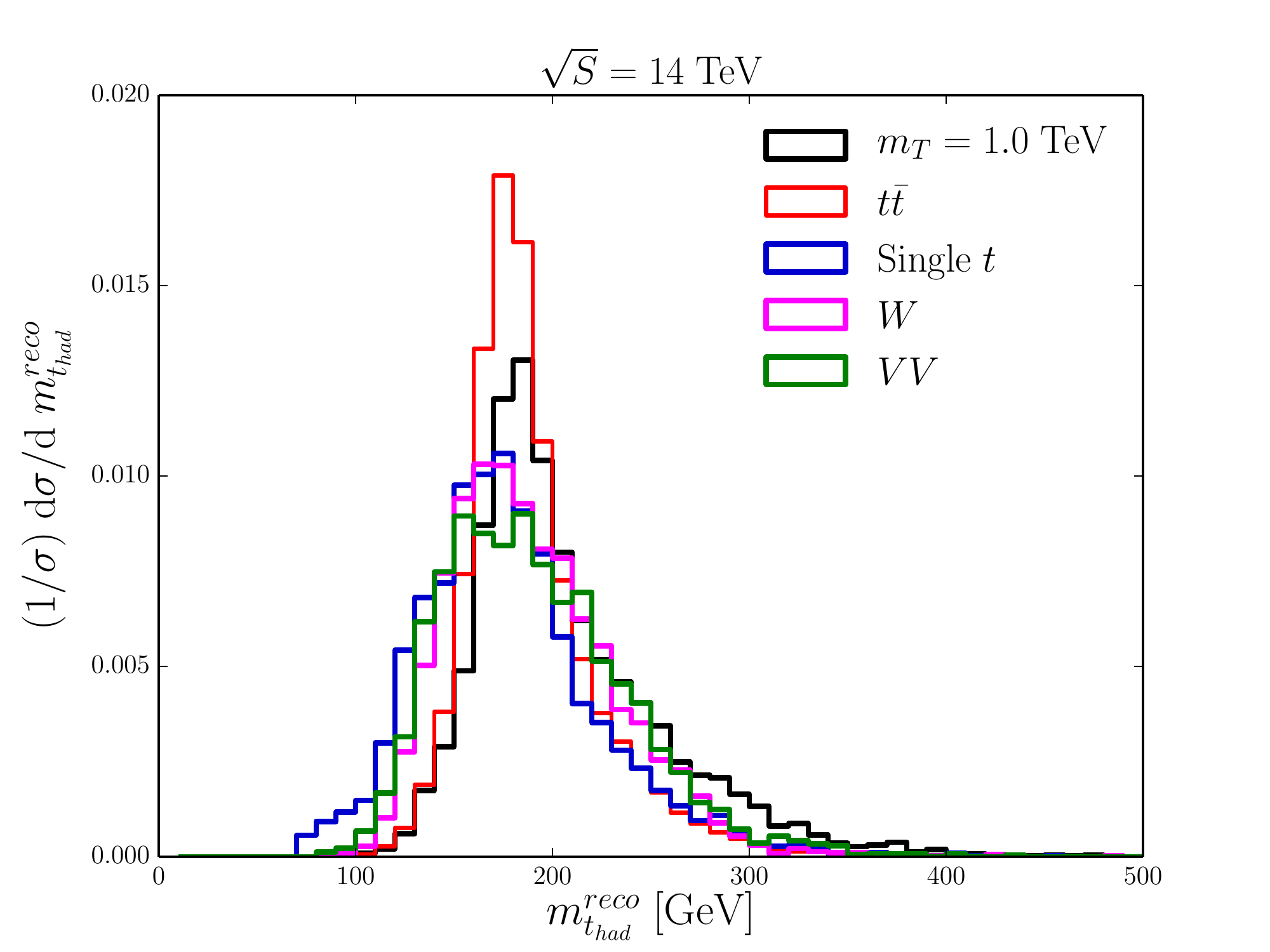} 
\includegraphics[width=0.49\textwidth,clip]{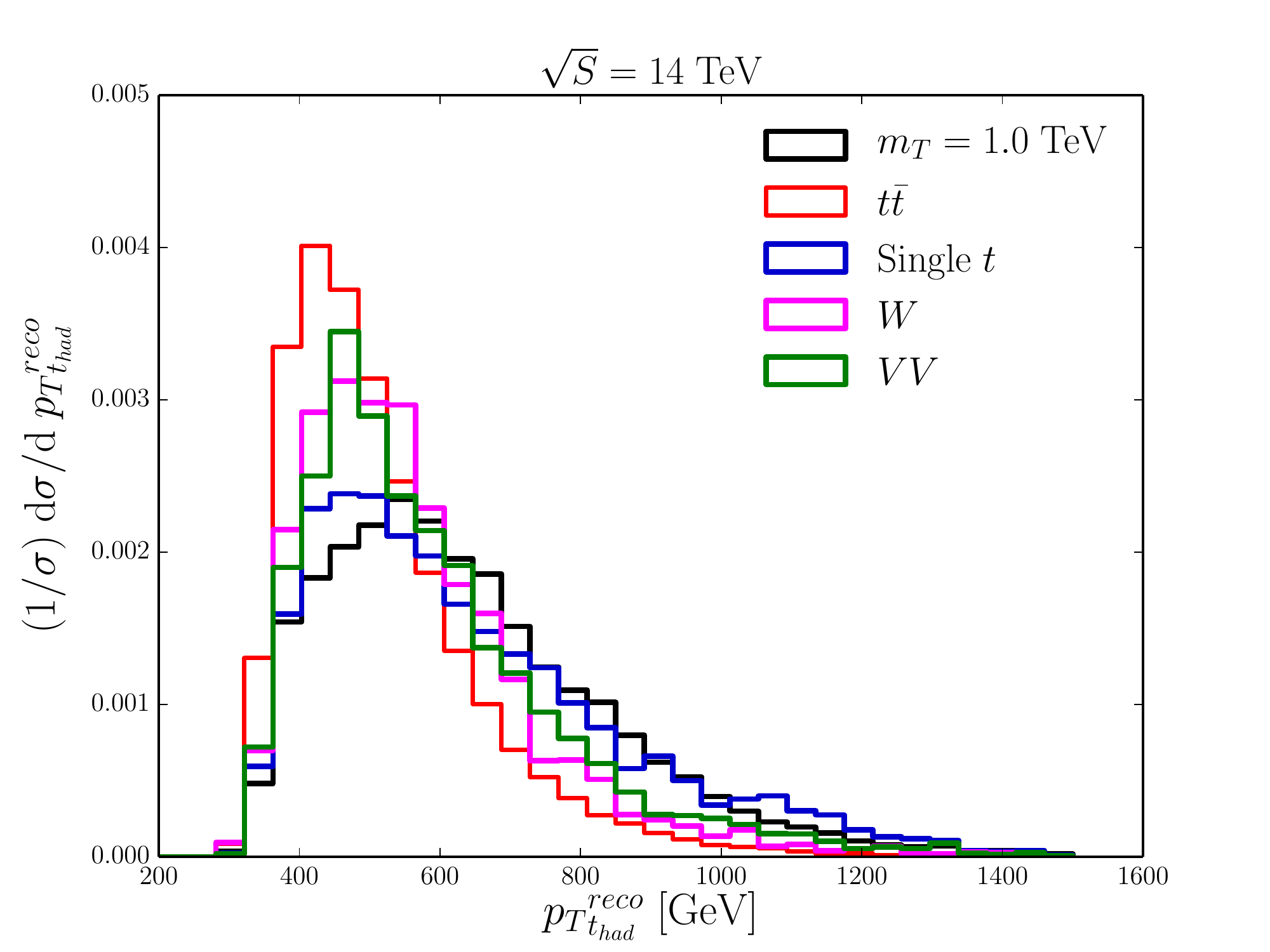}
\caption{The reconstructed invariant mass distribution of the top-tagged fat jet (left) and the corresponding $p_T$ distribution (right) in the $t \overline{t} g g$ channel for $m_{T} = 1.0\TeV$.}\label{fig:Top_gg}
\end{figure}
For a fat jet to be tagged as the hadronic top, we demand a leading order (LO) three-pronged top template overlap score 
\begin{eqnarray}
Ov_3^{had} > 0.6.\label{eq:fattop}
\end{eqnarray}
Fig.~\ref{fig:Top_gg} (left) shows the normalized invariant mass distributions of the top-tagged fat jet, $m_{t_{had}}^{reco}$, for $m_{T} = 1.0\TeV$ and both signal and background after reconstruction. Both the signal and $t\bar{t}$ background $m_{t_{had}}^{reco}$ distributions are highly peaked at the top mass $m_{t}=173$~GeV, while other backgrounds are slightly wider. Hence,  we apply the cut
\begin{eqnarray}
m_{t_{had}}^{reco}& > 145~{\rm GeV}  . \label{eq:TopM_cut}
\end{eqnarray}     
The corresponding $p_T$ distribution of the top-tagged fat jet is displayed in the right panel of Fig.~\ref{fig:Top_gg}, which shows that the signal is harder than the background. We require exactly one top-tagged fat jet which passes the cuts in Eqs.~(\ref{eq:fattop}-\ref{eq:TopM_cut}):
\begin{eqnarray}
N_{t_{had}}=1 \label{eq:Ntop} \, .
\end{eqnarray}

Table~\ref{tab:ttgg_Cutflow_1000} shows the cumulative effects of cuts on signal and background rates. Relative to the basic cuts in Eqs.~(\ref{eq:ETmiss}-\ref{eq:basefatjet}), under the requirement of $N_{t_{had}}=1$, the signal efficiency is $50 \%$, while the major backgrounds  $t \overline{t}$ and single $t$ have efficiencies of $59 \%$ and $30 \%$, respectively. The $W$ and $VV$ backgrounds are cut down to $19 \%$ and $21 \%$, respectively, reducing the overall size of the background. Typically, requiring at least one $b$-tagged jet\footnote{The slim $r=0.4$ jets are classified into three categories where our heavy-flavor tagging algorithm iterates over all jets that are matched to $b$-hadrons or $c$-hadrons. If a $b$-hadron ($c$-hadron) is found inside, it is classified as a $b$-jet ($c$-jet). The remaining unmatched jets are called light-jets. Each jet candidate is further multiplied by a tag-rate~\cite{ATL-PHYS-PUB-2016-026}, where we apply a flat $b$-tag rate of $\epsilon_{b \rightarrow b} = 0.7$ and a mis-tag rate that a $c$-jet (light-jet) is misidentified as a $b$-jet of $\epsilon_{c \rightarrow b} = 0.2$ ($\epsilon_{j \rightarrow b} = 0.01)$. For a $r= 1.0$ fat jet to be $b$-tagged, on the other hand, we require that a $b$-tagged $r = 0.4$ jet is found inside a fat jet. To take into account the case where more than one $b$-jet might land inside a fat jet, we reweight a $b$-tagging efficiency depending on a $b$-tagging scheme described in Ref.~\cite{Backovic:2015bca}.} in the top-tagged fat jet significantly improves the purity of the signal, suppresses non-resonant QCD backgrounds, and helps reduce the systematic uncertainty. However, since our dominant background is $t \overline{t} + \rm{jets}$, Table~\ref{tab:ttgg_Cutflow_1000} shows that the $b$-tagging merely degrades our final significance. Therefore we choose not to apply the $b$-tagging in our final results. 
\begin{table*}[t]
\begin{center}
\setlength{\tabcolsep}{1.5mm}
\renewcommand{\arraystretch}{1.1}
\scalebox{0.85}{
\centering
\begin{tabular}{|c|ccccc|c|c|}
\hline
  $t \overline{t} g g$ channel                                      &  Signal [fb]  & $t \overline{t}$ [fb]    & Single $t$ [fb]         & $W$ [fb]                   &  $VV$ [fb]                & $\sigma_{dis}$  & $\sigma_{excl}$ \\ \hline \hline
Basic cuts	                                                                  & 2.8              & $1.1 \times 10^{3}$  & $2.6 \times 10^{3}$ & $2.1 \times 10^{3}$  & 68                             & 2.0                  & 2.0     \\ 
$N_{t_{had}}=1$                                                        & 1.4              & 650                           & 790                          & 390                          & 14                             & 1.8                  &1.8      \\ 
$N_{t_{lep}}=1$                                                         & 0.60            & 140                           & 51                            & 28                            & 1.6                            & 2.2                  & 2.2    \\ 
{$p^{reco}_{T, \{ g_1, g_2 \}} > \{250, 150\}$ GeV }   & 0.35            & 9.2                         & 4.6                          & 2.5                         & 0.19                          & 4.8                  & 4.8 \\ 
$ H^{reco}_T > 1600$ GeV                                        & 0.29           & 4.9                          & 3.4                          & 1.6                         & 0.12                           & 5.1                  & 5.0 \\ 
{$ 750 < m^{reco}_{T_{1,2}} < 1100\ $ GeV}             & 0.16            & 0.84                         & 0.62                          & 0.23                          & 0.017                         & 6.7                  & 6.6 \\ \hline \hline
$b$-tag on $t_{\rm had}$                                           &0.10            &0.51                         &0.29                           &$5.6 \times 10^{-3}$   &$1.0 \times 10^{-3}$  & 5.9                  &5.8 \\ \hline \hline 
$b$-tag on $t_{\rm lep}$                                            &0.10            &0.49                         &0.21                           &0.016                          &$1.7 \times 10^{-4}$  & 6.4                  &6.3 \\ \hline \hline 
$b$-tag on $t_{\rm had} \; \& \;t_{\rm lep}$                & 0.061         &0.30                         &0.084                         &$5.1 \times 10^{-4}$   &$1.0 \times 10^{-5}$  & 5.3                  &5.2 \\ \hline  
\end{tabular}}
\end{center}
\caption{A cumulative cut-flow table showing the signal and SM background cross sections in the $t \overline{t} g g$ channel for $m_{T} = 1.0$ TeV. The significances $\sigma_{dis}$ and $\sigma_{excl}$ are calculated based on the likelihood-ratio methods defined in Eq. (\ref{Eq:SigDis}) and Eq. (\ref{Eq:SigExc}) respectively for a given luminosity of $3$ ab$^{-1}$.  The summary of the background simulations can be found in Table~\ref{tab:TotalBackG}. 
}
\label{tab:ttgg_Cutflow_1000}
\end{table*}

We now turn to the boosted leptonic top, $t_{lep}$, reconstruction~\cite{Backovic:2013bga} within the TOM framework.  The set of three-pronged templates used to tag the hadronically decaying top is also used to tag the leptonically decaying boosted top.  The overlap $Ov_3^{lep}$, where $lep$ denotes the leptonic top, is calculated using the four-momentum of a jet, the four-momentum of a lepton, and the missing transverse momentum ($\vec{\slashed{P}}_T$) \footnote{In the events we are considering, the only source of missing transverse momentum is the neutrino from the leptonically-decaying top.}.  For the leptonically decaying top quark,  there is missing longitudinal momentum from the neutrino that cannot be simply reconstructed since the initial state longitudinal momentum is unknown.  Hence, the full angular separation $\Delta R$ between the template and $\vec{\slashed{P}}_T$ cannot be determined, and the azimuthal distance $\Delta \phi$ between the template and $\vec{\slashed{P}}_T$ must be used to calculate $Ov_3^{lep}$.  Hence, in general the precise truth momentum of the top is not reconstructed.  However, the addition of $Ov_3^{lep}$ to our analysis still proves to be useful. i) We identify the lepton-jet pair originating from the leptonically decaying top quark as the pair that maximizes $Ov_3^{lep}$.  After this selection, for $85\%$ of the signal events, a $b$-hadron is found inside the selected jet as expected in a top quark decay. Therefore, this selection helps to resolve the combinatorial problem of determining which jet originates from the leptonically decaying top quark without the need for $b$-tagging.  This is useful for reconstructing top partner masses while maintaining signal efficiency. ii) It can reject the background events efficiently and boost signal sensitivity.

In what follows, we will demonstrate how the boosted $t_{lep}$ reconstruction works. We require at least one slim jet that is isolated from the hadronic top-tagged fat jet,  has $p_T^j > 30 \GeV$ and $|\eta^j| < 2.5$,  and meets the endpoint criterion 
\begin{eqnarray}
m_{j \ell} < 153.2  \GeV ,    \label{eq:endpoint}
\end{eqnarray}
where $m_{j \ell}$ is the invariant mass of the lepton-jet pair. We calculate the $Ov_3^{lep}$ score for each slim jet that passes these criteria, as described above.  
For a lepton-jet pair to be considered as decay products of the leptonic top, we demand
\begin{eqnarray}
Ov_3^{lep} > 0.5 .    \label{eq:lepOV}
\end{eqnarray}
The momentum of the corresponding (matched) three-pronged templates are used to reconstruct the four momentum of the $t_{lep}$, which in turn will be used to reconstruct the top partners.   We require exactly one $t_{lep}$ passing the cut in Eq.~(\ref{eq:lepOV}):
\begin{eqnarray}
N_{t_{lep}}=1 .  \label{eq:NtopLep}
\end{eqnarray}
Table~\ref{tab:ttgg_Cutflow_1000} shows that relative to the $N_{t_{had}}=1$ cut in Eq.~(\ref{eq:Ntop}), under the requirement of $N_{t_{lep}}=1$ the signal efficiency is $43 \%$, while the major backgrounds  $t \overline{t}$ and single $t$ have efficiencies of $22 \%$ and $6.5 \%$, respectively. The efficiencies of $W$ and $VV$ backgrounds are $7.2 \%$ and $11 \%$, respectively, greatly suppressing the overall size of backgrounds.
\begin{figure}[tb]
\centering
\includegraphics[width=0.49\textwidth,clip]{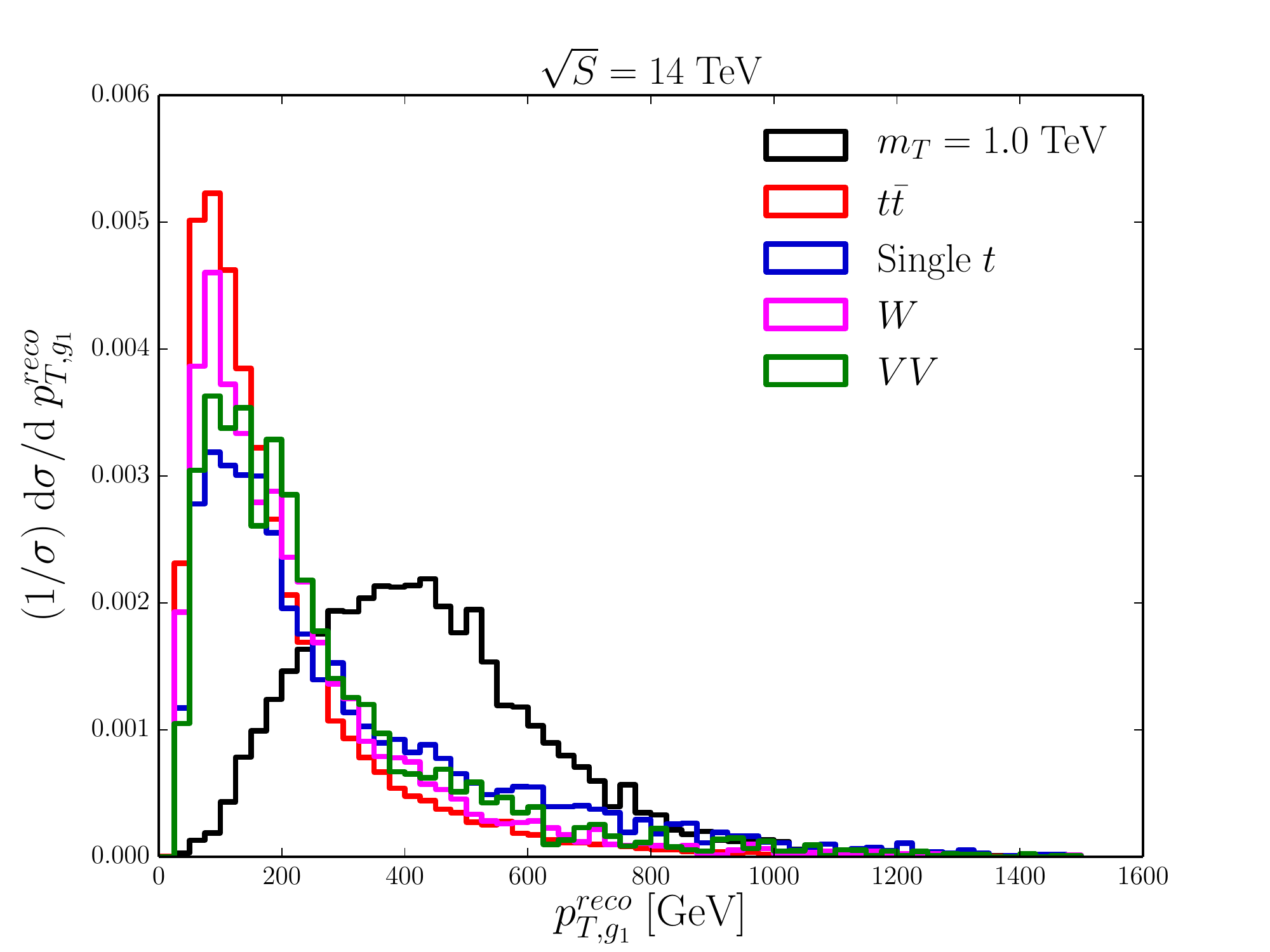}
\includegraphics[width=0.49\textwidth,clip]{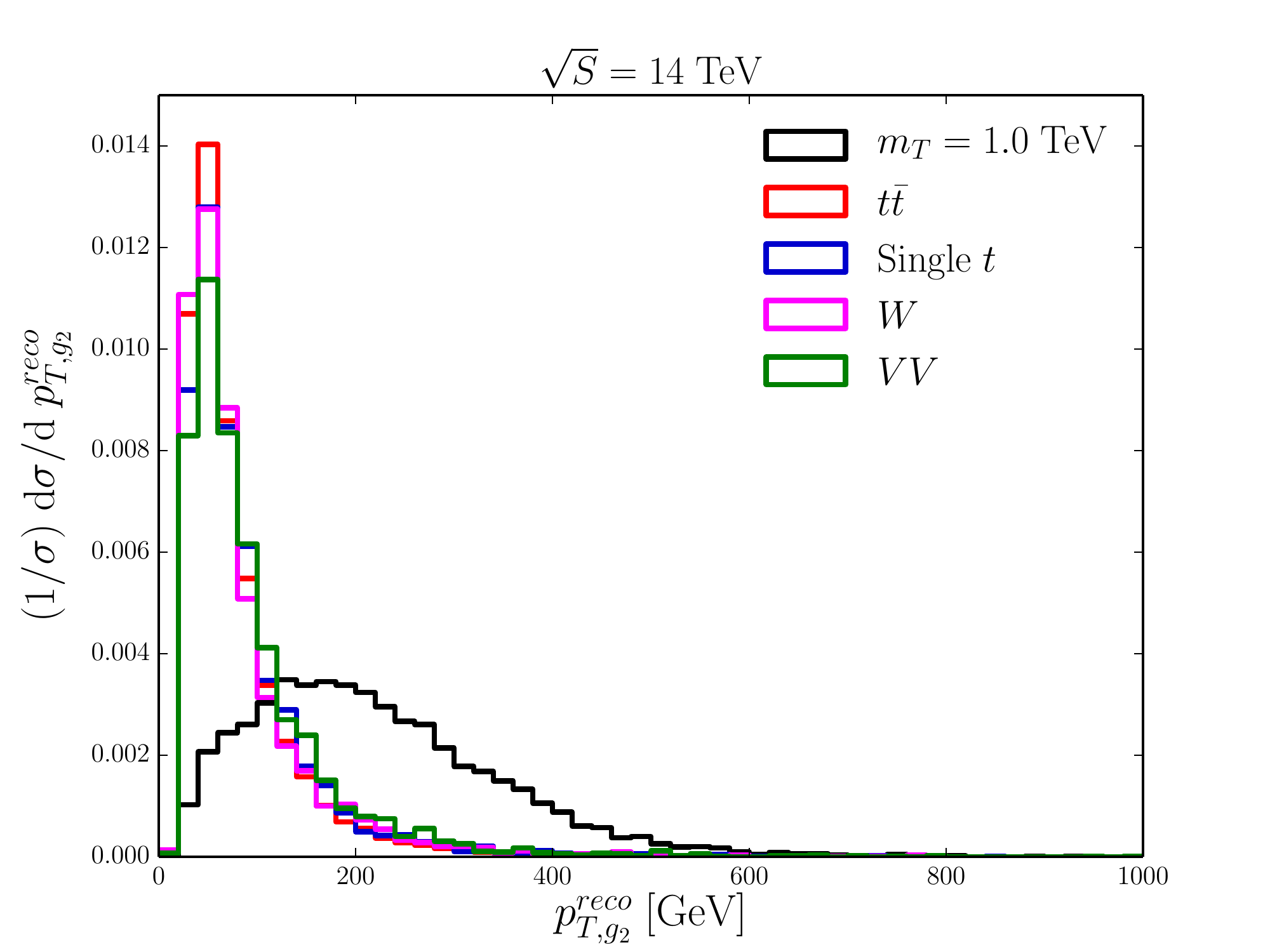} 
\includegraphics[width=0.49\textwidth,clip]{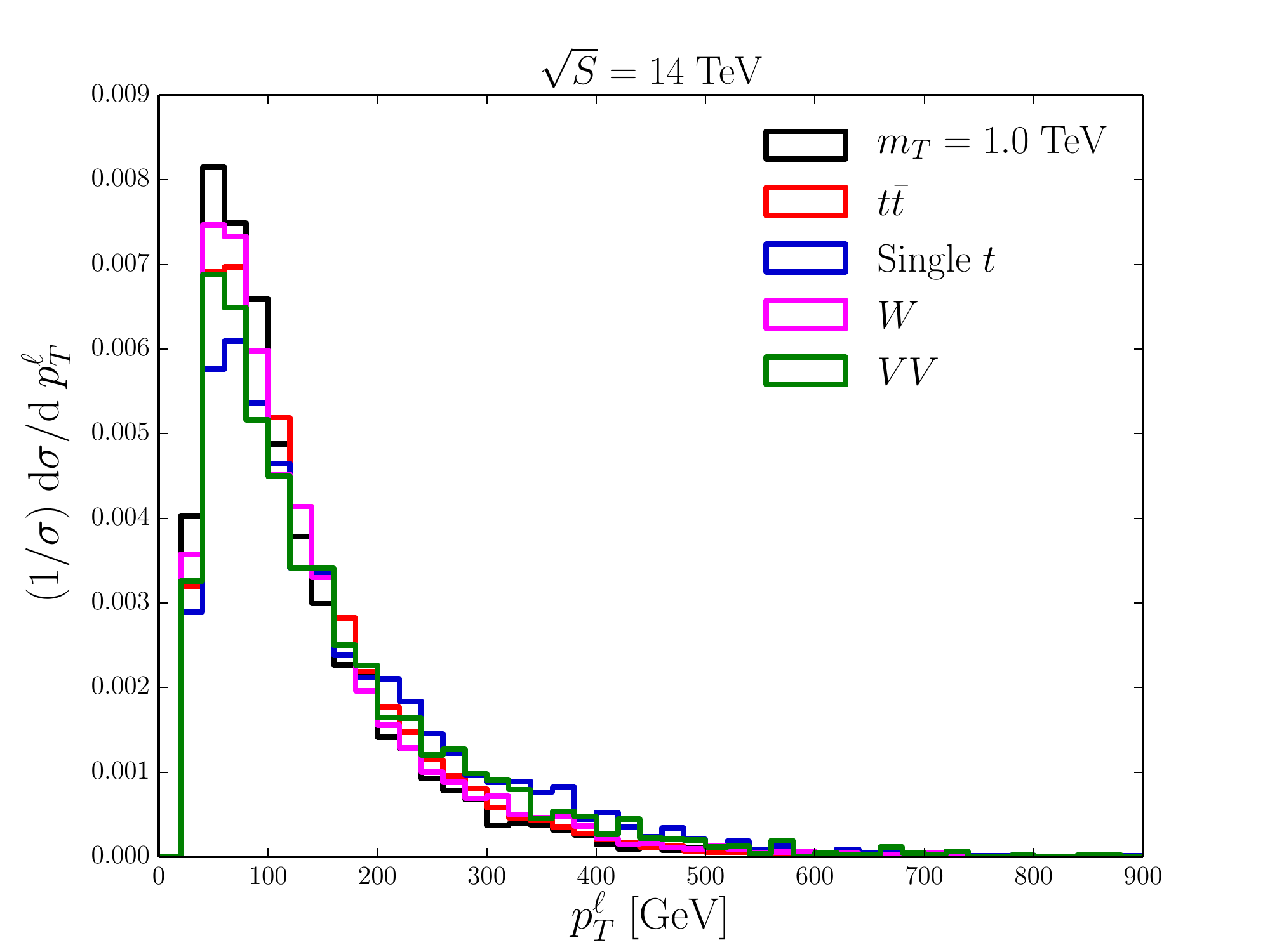}
\includegraphics[width=0.49\textwidth,clip]{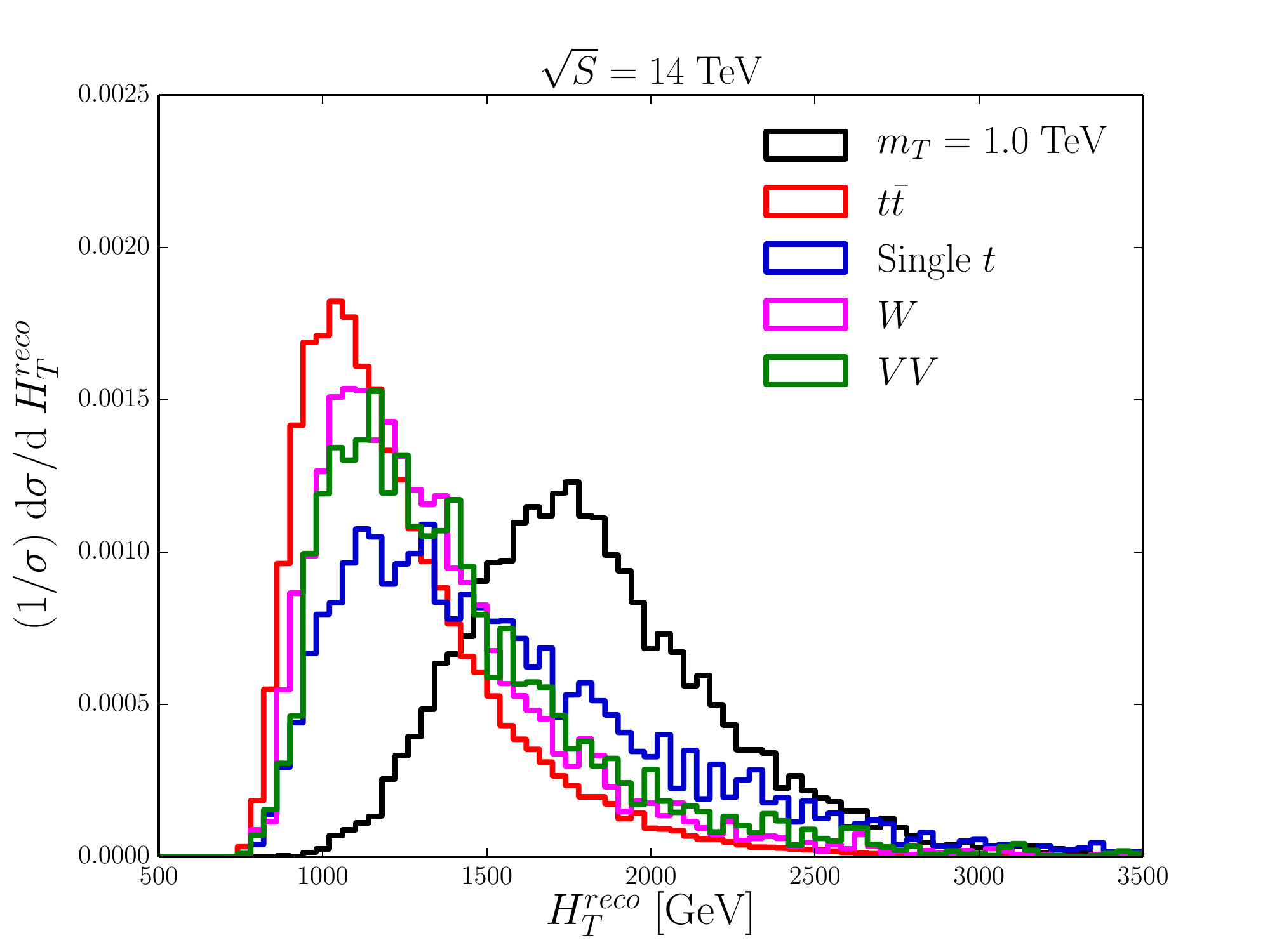}
\caption{The $p_T$ distributions of the first (top-left) and second (top-right) hardest slim jets not associated with $t_{lep}$ or $t_{had}$, and (bottom-left) the isolated lepton, in the $t \overline{t} g g$ channel for $m_{T} = 1.0\TeV$.
The scalar sum, $H_T^{reco}$ in Eq.~(\ref{eq:HTreco}), of the transverse momenta of reconstructed hadronic and leptonic tops, and the two hardest slim jets is shown in the bottom-right panel.\label{fig:gg}}
\end{figure}

Additionally, since two hard gluons originate from the top partner decays, we require two additional slim jets not associated with the reconstructed tops with $p_T^j > 30 \GeV$ and $|\eta^j| < 2.5$ and well-separated from the top quarks by $\Delta R>1.4$. Fig. \ref{fig:gg} shows the $p_T$ distributions of the (top-left) first and (top-right) second hardest jets that are not part of the reconstructed $t_{had}$ or $t_{lep}$ for $m_{T} = 1.0\TeV$.  The lepton $p_T$ distribution is shown in the bottom-left plot of Fig.~\ref{fig:gg}.  As can be clearly seen, the signal jets are much harder than the background jets.  Hence, we place the further cuts on the two hardest jets not associated with $t_{had}$ or $t_{lep}$:
\begin{eqnarray}
p^{reco}_{T, g_1} > 250~{\rm GeV}\quad\quad{\rm and}\label{eq:gluonPT}\quad\quad
p^{reco}_{T, g_2} > 150~{\rm GeV}. \,
\end{eqnarray} 
As shown in Table~\ref{tab:ttgg_Cutflow_1000}, relative to the $N_{t_{lep}}$ cut of Eq.~(\ref{eq:NtopLep}), the signal efficiency is 58\% while the overall background efficiency is 7.5\%.  Hence, this is a key driver to overall background suppression.

\begin{figure}[tb]
\centering
\includegraphics[width=0.49\textwidth,clip]{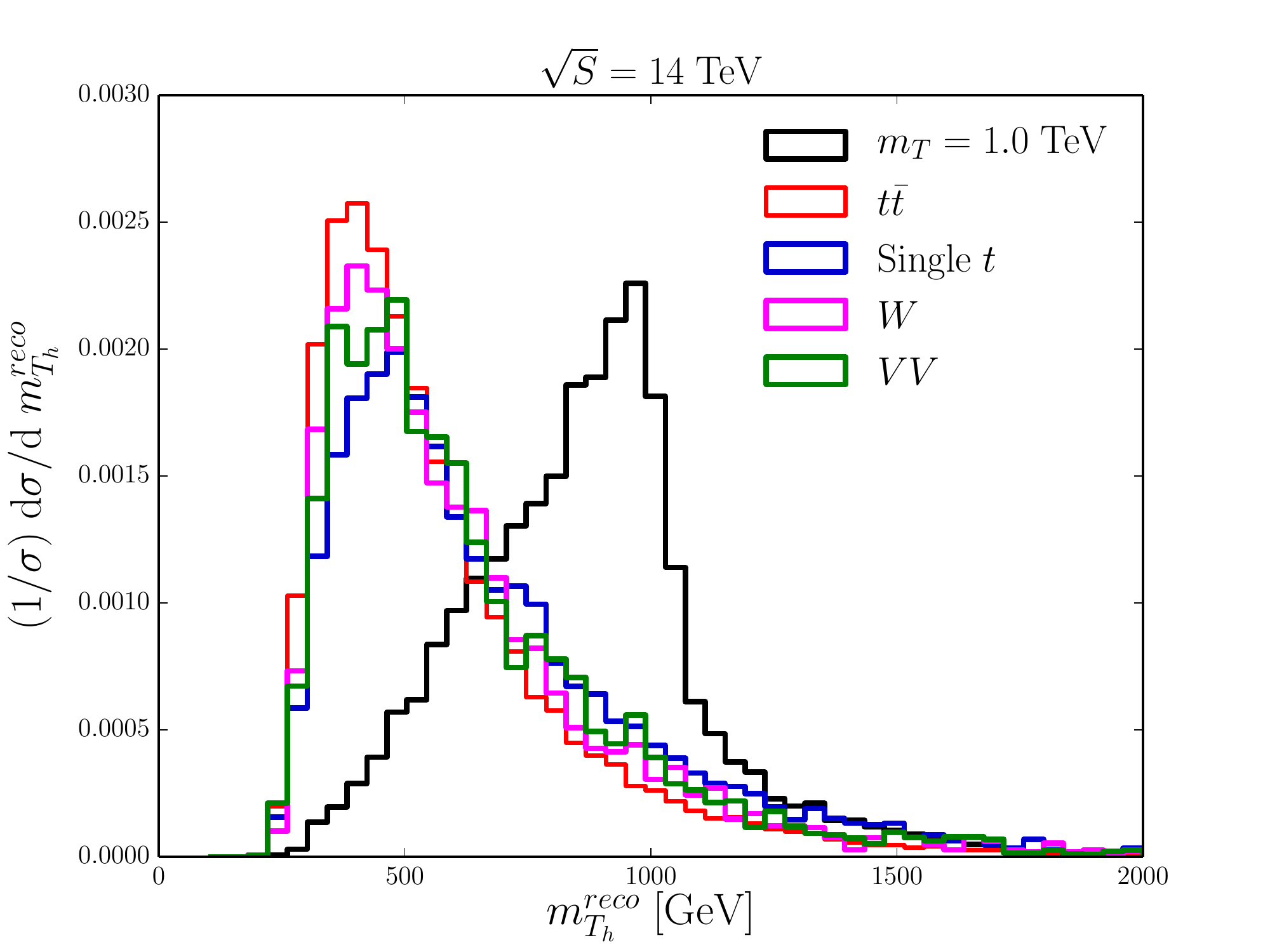}
\includegraphics[width=0.49\textwidth,clip]{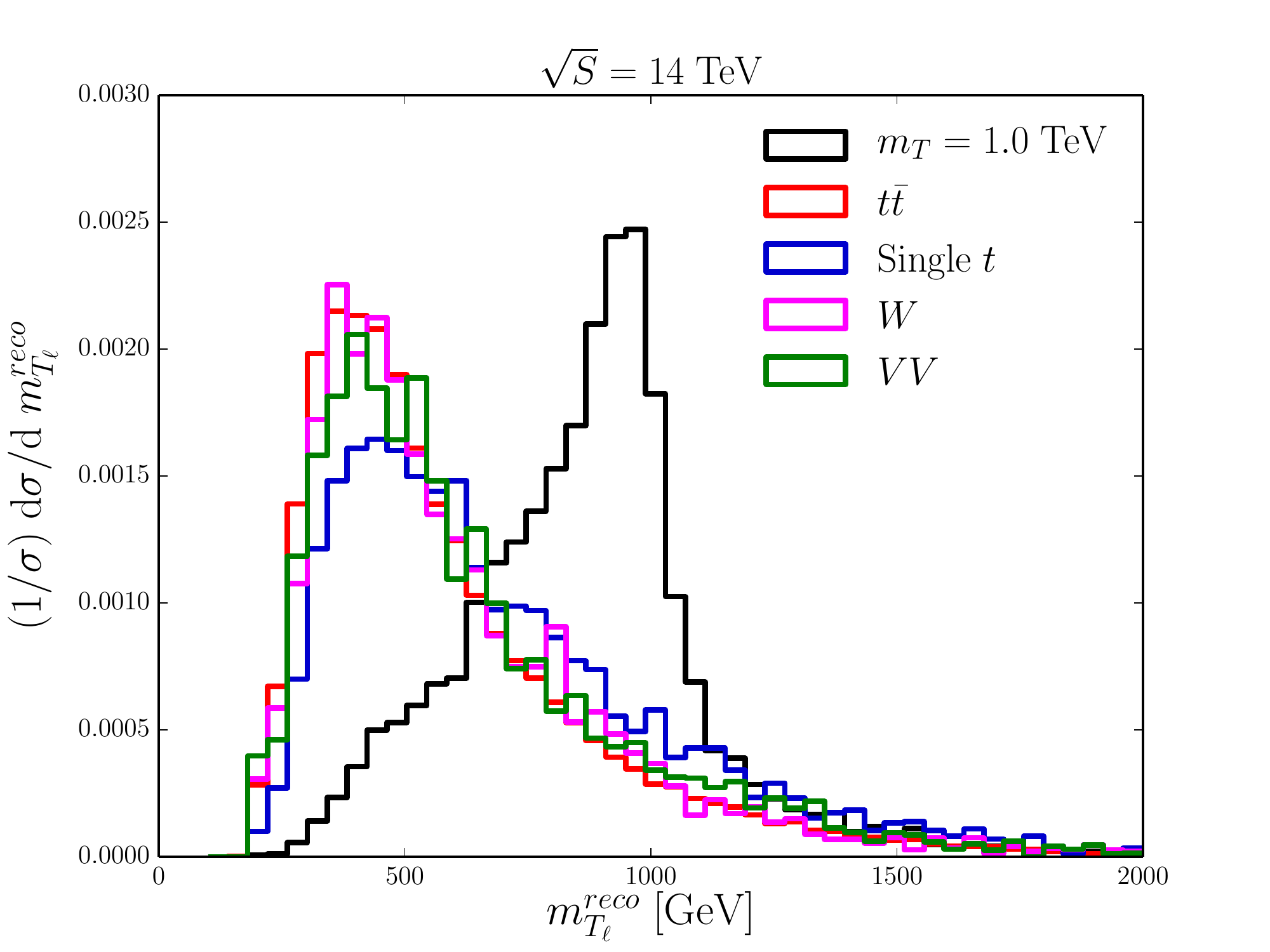} 
\caption{The distributions of reconstructed top partner invariant masses $m^{reco}_{T_{h}}$ (left) and $m^{reco}_{T_{\ell}}$ (right) in the $t \overline{t} g g$ channel for $m_{T} = 1.0\TeV$.\label{fig:MtprimeGG}}
\end{figure}

To further exploit the boosted phase space of signal events, we introduce the variable $H^{reco}_T$ defined as the scalar sum of the transverse momenta of the reconstructed hadronic top, the reconstructed leptonic top, and the first two hardest jets isolated from $t_{had}$ and $t_{lep}$: 
\bea
\label{eq:HTreco} 
	H^{reco}_T  =  p_{T,t_{had}}^{reco} + p_{T,t_{lep}}^{reco} + p^{reco}_{T, g_1} + p^{reco}_{T, g_2} \;.
\eea
The $H^{reco}_T$ is somewhat correlated with the cuts introduced in Eqs. (\ref{eq:basefatjet}) and (\ref{eq:gluonPT}), but allows us to directly control the total transverse energy of the reconstructed final states. The bottom-right of Fig.~\ref{fig:gg} shows the $H^{reco}_T$ distributions for both signal and background prior to applying the cuts in Eqs. (\ref{eq:basefatjet}) and (\ref{eq:gluonPT}). Again, the signal is clearly much harder than the background and for a higher significance we apply the cut
\begin{eqnarray}
H^{reco}_T > 1600~{\rm GeV} . \label{eq:HTCut}
\end{eqnarray}

With the top quark reconstruction and the two hardest jets, we can now reconstruct the top partners. After imposing the series of cuts in Eqs. (\ref{eq:Ntop}) and (\ref{eq:NtopLep}-\ref{eq:HTreco}), the phase space of the SM backgrounds is carved into the signal region. The only remaining information orthogonal between signal and background is the top partner invariant mass. While both hadronic and leptonic tops are fully reconstructed, it is not clear yet which combination of top quarks and additional hard slim jets originate from the same top partner decay. Of two possible combinations of top quarks and jets, we reconstruct the top partners by using the symmetry of the top partner decays and minimizing the asymmetry
\begin{equation}
\Delta_m  \equiv \mathrm{min} \left[ |  m^{reco}_{t_{had}  g_1}  - m^{reco}_{t_{lep} g_2} |, | m^{reco}_{t_{had} g_2}  - m^{reco}_{t_{lep} g_1} |\right] \,,  \label{eq:DeltaM}
\end{equation}
where $ m^{reco}_{t_{i} g_{j}} $ stands for the invariant mass of the pair $\{ t_{i} , g_{j}  \}$, $i=had,lep$ denotes either the hadronic or leptonic top, and $j=1,2$ indicates either the first or second hardest additional slim jets. The reconstructed top partners are identified as the pair of $\{t_{had},g_i\},\{t_{lep},g_j\}$, $i\neq j$, that minimize $\Delta_m$.  The resulting distributions of the reconstructed invariant masses of the hadronically decaying top partner ($m^{reco}_{T_h}$) and leptonically decaying top partner ($m_{T_\ell}^{reco}$) are shown in Fig.~\ref{fig:MtprimeGG} for $m_T=1.0$~TeV. Although they both display sizable lower tails, they peak at the truth-level top partner mass. Since the backgrounds are populated at much lower invariant mass, they can be separated by the cut
\begin{eqnarray}
750~{\rm GeV}<&m^{reco}_{T_{h,\ell}}&<1100~{\rm GeV} . \label{eq:MtprimeCut}
\end{eqnarray}
The effects of the mass window cut in Eq. (\ref{eq:MtprimeCut}) are shown in Table~\ref{tab:ttgg_Cutflow_1000}, where the dominant $t \overline{t}$ background is brought down to the same order of magnitude as the signal cross section. 

We summarize the cumulative effects of cuts on signal and background cross sections (in fb) in Table~\ref{tab:ttgg_Cutflow_1000}. To quantify the discovery reach of our signal at the LHC, we compute a significance ($\sigma_{dis}$) for discovery  using the likelihood-ratio method~\cite{Cowan:2010js}
\beq
  \sigma_{dis} \equiv
    \sqrt{-2\,\ln\bigg(\frac{L(B | Sig\!+\!B)}{L( Sig\!+\!B| Sig\!+\!B)}\bigg)}
  \;\;\;\;\; \text{with}\;\;\;
  L(x |n) =  \frac{x^{n}}{n !} e^{-x} \,,
\label{Eq:SigDis} \eeq
where $Sig$ and $B$ are the expected number of signal and background events, respectively. For a discovery we demand
\beq
	\sigma_{ dis} \geq 5.
\eeq
To set an exclusion limit on our signal, we compute a significance ($\sigma_{excl}$) for exclusion using a different likelihood-ratio 
\beq
  \sigma_{exc} \equiv
    \sqrt{-2\,\ln\bigg(\frac{L(Sig\!+\!B | B )}{L( B | B)}\bigg)} .
\label{Eq:SigExc} 
\eeq
For an exclusion we demand
\beq
	\sigma_{excl} \geq 2.
\eeq
 All significances $\sigma_{dis}$ and $\sigma_{excl}$ in Table~\ref{tab:ttgg_Cutflow_1000} are calculated for a given luminosity of $3$ ab$^{-1}$.

The seventh row of Table~\ref{tab:ttgg_Cutflow_1000} shows our final results for signal rates, background rates, and discovery and exclusion significances after all cuts in Eqs.~(\ref{eq:basethinjet}-\ref{eq:MtprimeCut}).  
The outlook for the $t \overline{t} g g$ channel is quite promising with a discovery significance of $\sigma_{dis} = 6.7$ at the high luminosity LHC for $m_{T} = 1.0$ TeV. The cornerstones of our search strategy are the boosted hadronic and leptonic top reconstructions, which enabled us to fully reconstruct top partner invariant masses. With the invariant mass cuts in Eq. (\ref{eq:MtprimeCut}), we find a factor of $1.3$ improvement in the discovery and exclusion significances relative to the cuts in Eqs.~(\ref{eq:basethinjet}-\ref{eq:HTCut}). The effectiveness of the reconstructed top partner mass cuts rapidly increases as we probe top partner masses higher than 1 TeV, since the backgrounds are populated at a much lower invariant mass region. On the other hand, the cuts on the additional hard slim jet transverse momenta in Eq (\ref{eq:gluonPT}) deliver the biggest improvement by increasing the significances by a factor of $2.2$ relative to the cuts in Eqs.~(\ref{eq:basethinjet}-\ref{eq:NtopLep}). The effect is attributed to the fact that jet activity in the $t \overline{t}$ background originates from QCD and is generally softer than the $p_T$ cuts in Eq. (\ref{eq:gluonPT}). 

In the last three rows of Table~\ref{tab:ttgg_Cutflow_1000} we show the effects of $b$-tagging.  As mentioned earlier, we find that $b$-tagging on the hadronic and leptonic tops decreases the final signal significance. The main $t\bar{t}$ background and signal are suppressed by the same $b$-tagging efficiency since they both have the same number of final state $b$-jets.  Since after all cuts the rate of $t\bar{t}$ is still five times that of the signal, decreasing both cross sections at the same rate suppresses the overall significance. It should be emphasized, however, that $b$-tagging proves to be effective in suppressing the other backgrounds.

\subsection{$t \overline{t} g \gamma$ Decay Channel}\label{sec:analysis2}
Although extensive searches have been carried out for top partner pair production, to our knowledge no previous study has investigated the $t \overline{t} g \gamma$ channel and this will be the first paper to assess the discovery potential of this final state.  Due to the presence of the hard photon, $t \overline{t} g \gamma$ is much cleaner than $t \overline{t} g g$ and has less contamination from SM backgrounds. On top of that, since the photon can be remarkably well measured, the resolutions of reconstructed $T$ invariant masses will be much better than in the $t \overline{t} g g$ channel.  This will give us a better handle for extracting signal from background. In this section, we will repeat a similar analysis as that presented in Section \ref{sec:analysis1} with minor modifications to maximize the use of the isolated photon.  We will demonstrate that the $t \overline{t} g \gamma$ channel outperforms $t \overline{t} g g$ channel in a wide range of parameter space.

The dominant background is $t \overline{t} \gamma + \rm{jet}$ matched with up to one additional jet where the tops are decayed semi-leptonically. The next important background is $t \gamma$ process including $tW \gamma$ and $tq \gamma$, where $q$ is a light quark or a $b$-quark.  The $tW \gamma$ background is generated with up to two additional jets where one $W$ decays leptonically while the other decays hadronically. The $tq \gamma$ process is generated with up to three additional jets and we only consider a top quark which decays leptonically. The sub-leading background includes $W \gamma $ matched with up to three additional jets where  the $W$ is decayed leptonically. The other non-significant backgrounds include $W W \gamma$ matched with up to two additional jets where one $W$ decays leptonically and the other hadronically. The $W Z \gamma$ background is matched with up to two additional jets where the $W$ and $Z$ are decayed leptonically and hadronically, respectively. The background events are simulated at a $\sqrt{S} = 14$ TeV in the same set-up described in Section \ref{sec:analysis1}. The generation-level cuts in Eqs.~(\ref{eq:basepartons}-\ref{eq:basephoton}) are applied, and Table \ref{tab:TotalBackA} summarizes the background simulations. All background events are showered, hadronized and smeared accordingly.
 
The other important backgrounds are due to jets faking photons. We have implemented in our background analysis the jet-to-photon misidentification rate as a function of $p_T^j$ following Ref. \cite{ATL-PHYS-PUB-2016-026,Goncalves:2018qas}.  It has been verified that the jets faking photon backgrounds are not relevant.  This is because our photons are very energetic and the corresponding fake rate is very small at an order of $\lesssim 10^{-4}$.
\begin{table}[t!]
\begin{center}
\scalebox{1.0}{
\begin{tabular}{|c|c|c|c|c|}
\hline
           Abbreviations                   & Backgrounds                               & Matching     &  $\sigma \cdot {\rm BR(fb)}$    \\ \hline	
  $t \bar{t} \gamma$                    &  $t \bar{t} + \gamma + \rm{jet}$ &   4-flavor      &  $  1.0 \; \rm fb $                        \\ \hline
 \multirow{2}{*}{$t \gamma$}      & $t W + \gamma + \rm{jets}$         &    5-flavor     & $  1.9 \; \rm fb $                      \\
                                                    &$t + \gamma + \rm{jets}$              &    4-flavor    &  	$ 0.085 \; \rm fb$               \\ \hline
  $W \gamma$                            &$W + \gamma + \rm{jets}$           &     5-flavor     &   $ 5.4     \; \rm fb $                \\ \hline  
  \multirow{2}{*}{$VV \gamma$} & $W W + \gamma + \rm{jets}$      &    4-flavor      &   $ 0.17  \; \rm fb$             \\ 
                                                   &$W Z + \gamma + \rm{jets}$        &    4-flavor      &    $ 0.057  \; \rm fb$                   \\

\hline		
\end{tabular}}
\end{center}
\caption{The summary of the SM backgrounds relevant to the $t \overline{t} g \gamma$ channel and their cross section after generation level cuts Eqs.~(\ref{eq:basepartons}-\ref{eq:basephoton}).  Matching refers to the either the 4-flavor or 5-flavor MLM matching~\cite{MLM}. The last column $\sigma \cdot \rm BR$ denotes the production cross section (in fb) times branching ratios including the top, $W$, and $Z$ decays.}
\label{tab:TotalBackA}
\end{table}

Basic selection cuts on leptons and jets are the same as those in Eqs.~(\ref{eq:ETmiss}-\ref{eq:basefatjet}).  We additionally require exactly one isolated photon with
\begin{eqnarray}
 p^{\Sigma}_T / p_T^{\gamma} < 0.1,\label{eq:isophoton}
\end{eqnarray}
where $p^{\Sigma}_T$ is the scalar sum of transverse momenta of final state particles (excluding the photon) in a cone of size $\Delta R = 0.4$. 
The set of cuts in Eqs.~(\ref{eq:ETmiss}-\ref{eq:basefatjet}) and (\ref{eq:isophoton}) defines our basic cuts of the $t \overline{t} g \gamma$ channel.

\begin{figure}[t!]
\begin{center}
\includegraphics[width=0.49\textwidth,clip]{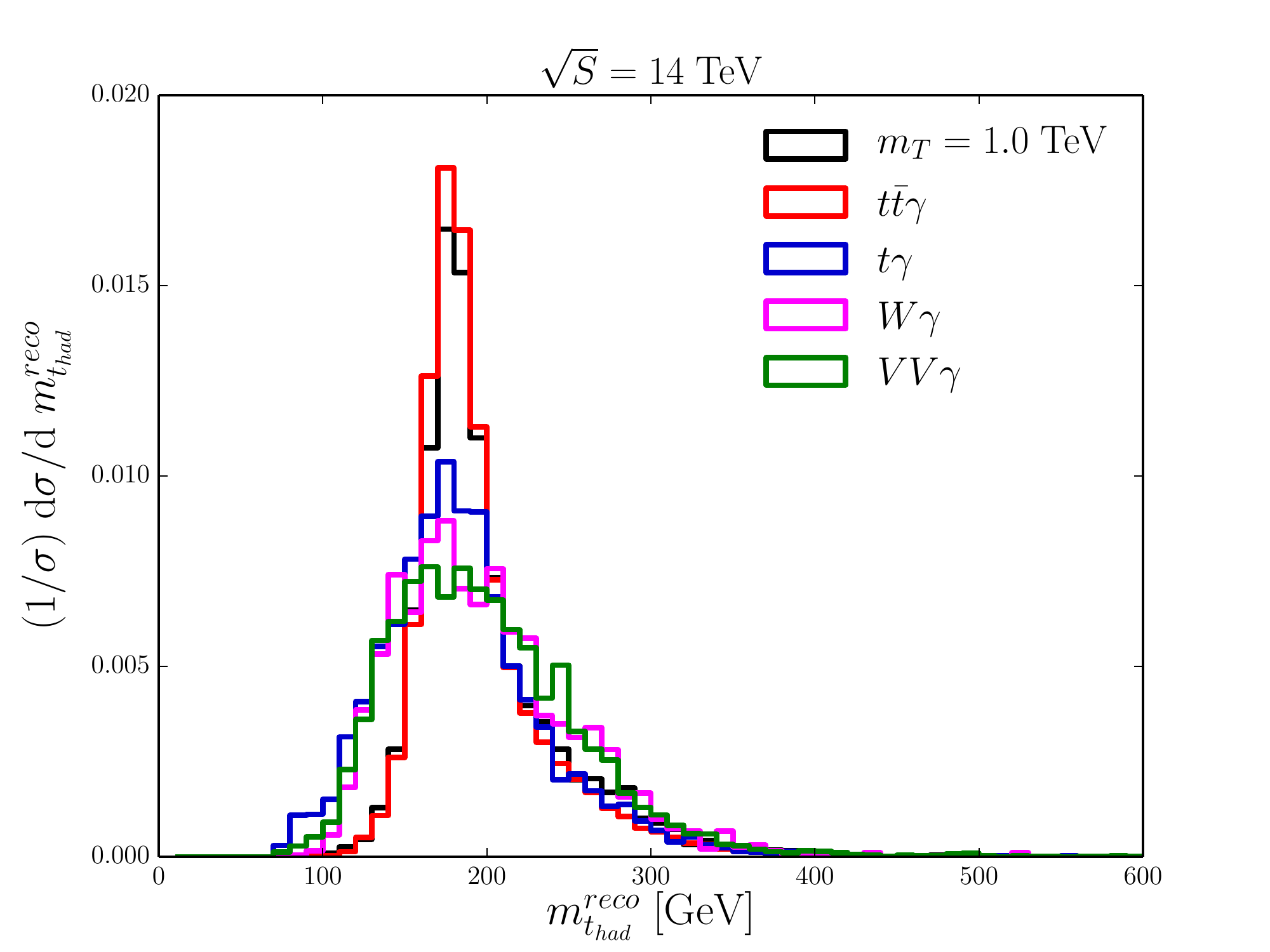}  
\includegraphics[width=0.49\textwidth,clip]{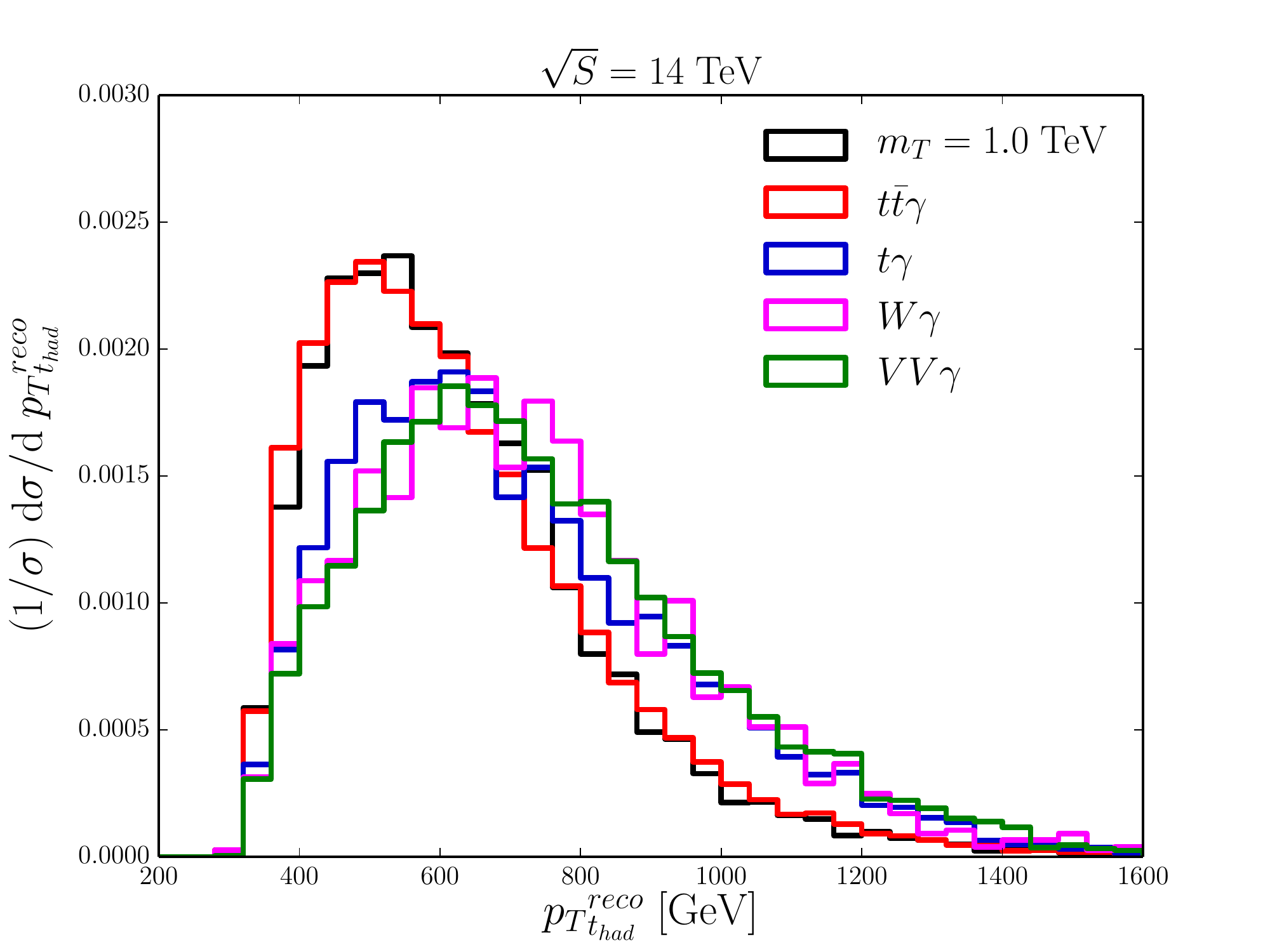}
\caption{The reconstructed invariant mass distribution of the top-tagged fat jet (left) and the corresponding $p_T$ distribution (right) in the $t \overline{t} g \gamma$ channel for $m_{T} = 1.0\TeV$.\label{fig:Top_ga}}
\end{center}
\end{figure}

We require exactly one top-tagged fat jet which passes the cuts in Eqs.~(\ref{eq:fattop}-\ref{eq:TopM_cut}):
\begin{eqnarray}
N_{t_{had}}=1  , \label{eq:Ntop_ga}
\end{eqnarray}
and exactly one boosted leptonic top passing the cut in Eq.~(\ref{eq:lepOV}):
\begin{eqnarray}
N_{t_{lep}}=1 .  \label{eq:NtopLep_ga}
\end{eqnarray}
Distributions of the reconstructed invariant mass of the top-tagged fat jet and the corresponding $p_T$ are displayed in Fig. \ref{fig:Top_ga}.

\begin{figure}[tb]
\centering
\includegraphics[width=0.49\textwidth,clip]{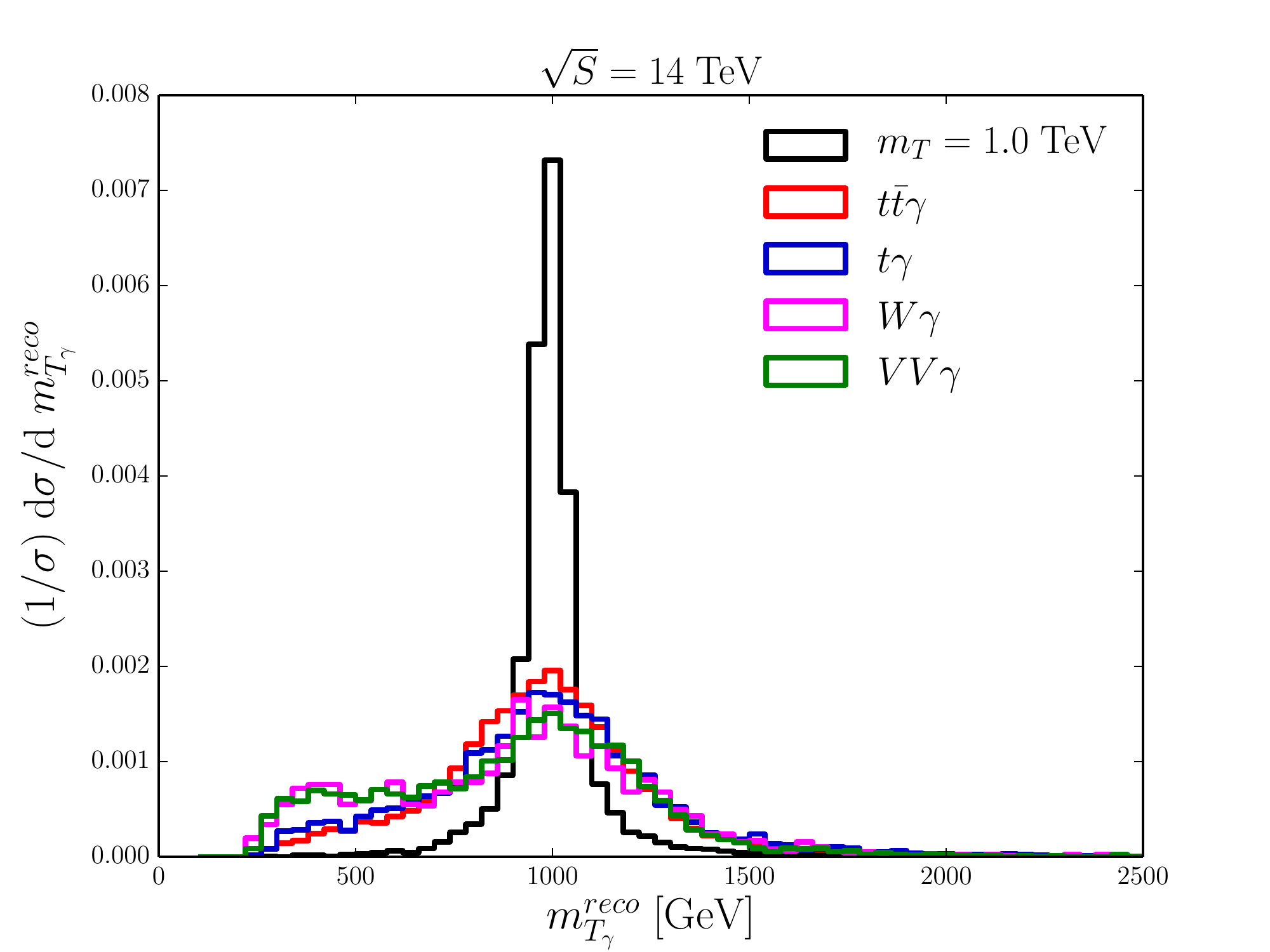}
\includegraphics[width=0.49\textwidth,clip]{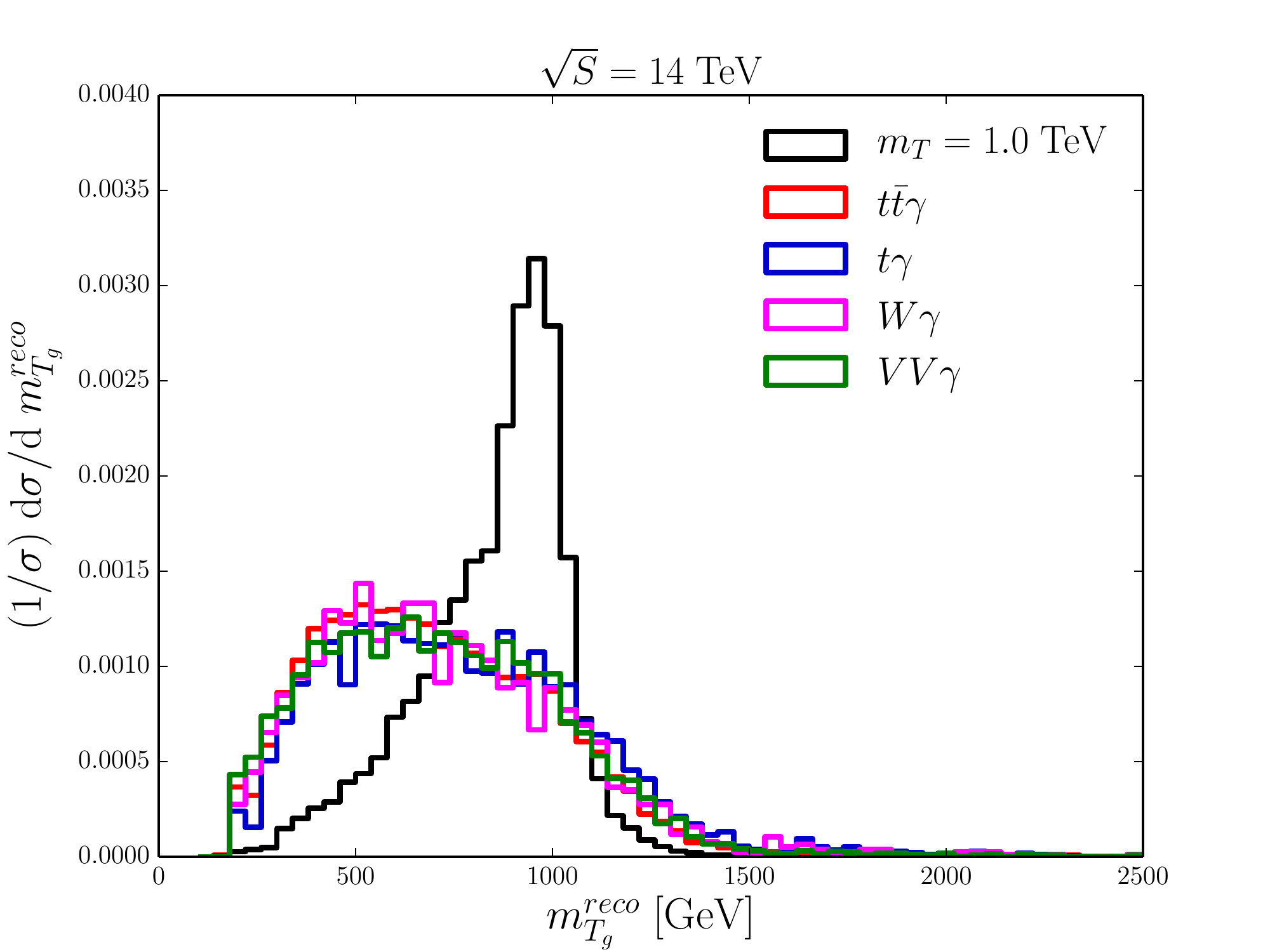}
\caption{The distributions of reconstructed top partner invariant masses $m^{reco}_{T_{\gamma}}$ (left) and $m^{reco}_{T_{g}}$ (right) in the $t \overline{t} g \gamma$ channel for $m_{T} = 1.0\TeV$. 
\label{fig:MtprimeGA}}
\end{figure}

We also demand at least one slim jet with $p_T^j > 30 \GeV$ and $|\eta^j| < 2.5$ that is well separated from the reconstructed $t_{had}$ and $t_{lep}$ by $\Delta R > 1.4$. To identify which jet originates from a top partner decay and reconstruct each top partner, we utilize the asymmetry of 
\begin{equation}
\Delta_m  \equiv   |  m^{reco}_{t_{k}  g_i}  - m^{truth}_T |^2 + |  m^{reco}_{t_{k^\prime}  \gamma}  - m^{truth}_T |^2  \label{eq:DeltaM_ga}
\end{equation}
 where we iterate over all the well-separated slim jets $g_i$, $k, \, k^\prime=had,lep$ ($k\neq k^\prime$) indicates either the hadronic or leptonic reconstructed top, and $m^{truth}_T$ is the truth (hypothesized) top partner mass.  The combination of $\{t_k,g_i\}$ and $\{t_{k'},\gamma\}$ that minimizes $\Delta_m$ in Eq.~(\ref{eq:DeltaM_ga}) is identified as the reconstructed top partners.  The reconstructed invariant mass of the top partner that decays into $t\gamma$ or $tg$ is denoted by $m^{reco}_{T_{\gamma}}$ or $m^{reco}_{T_{g}}$, respectively. As shown in Fig.~\ref{fig:MtprimeGA}, the distribution of $m^{reco}_{T_{\gamma}}$ is much narrower than $m^{reco}_{T_{g}}$ since the photon can be well measured and is less sensitive to the detector smearing. Since the background distributions of the reconstructed top partner masses are much broader, they can be separated by the cuts
 \begin{eqnarray}
900<&m^{reco}_{T_{\gamma}}&<1100~{\rm GeV} \quad\quad{\rm and}  \quad \quad 700< m^{reco}_{T_{g}}<1100~{\rm GeV}\,  . \label{eq:MtprimeCut_ga}
\end{eqnarray}

\begin{figure}[tb]
\centering
\includegraphics[width=0.49\textwidth,clip]{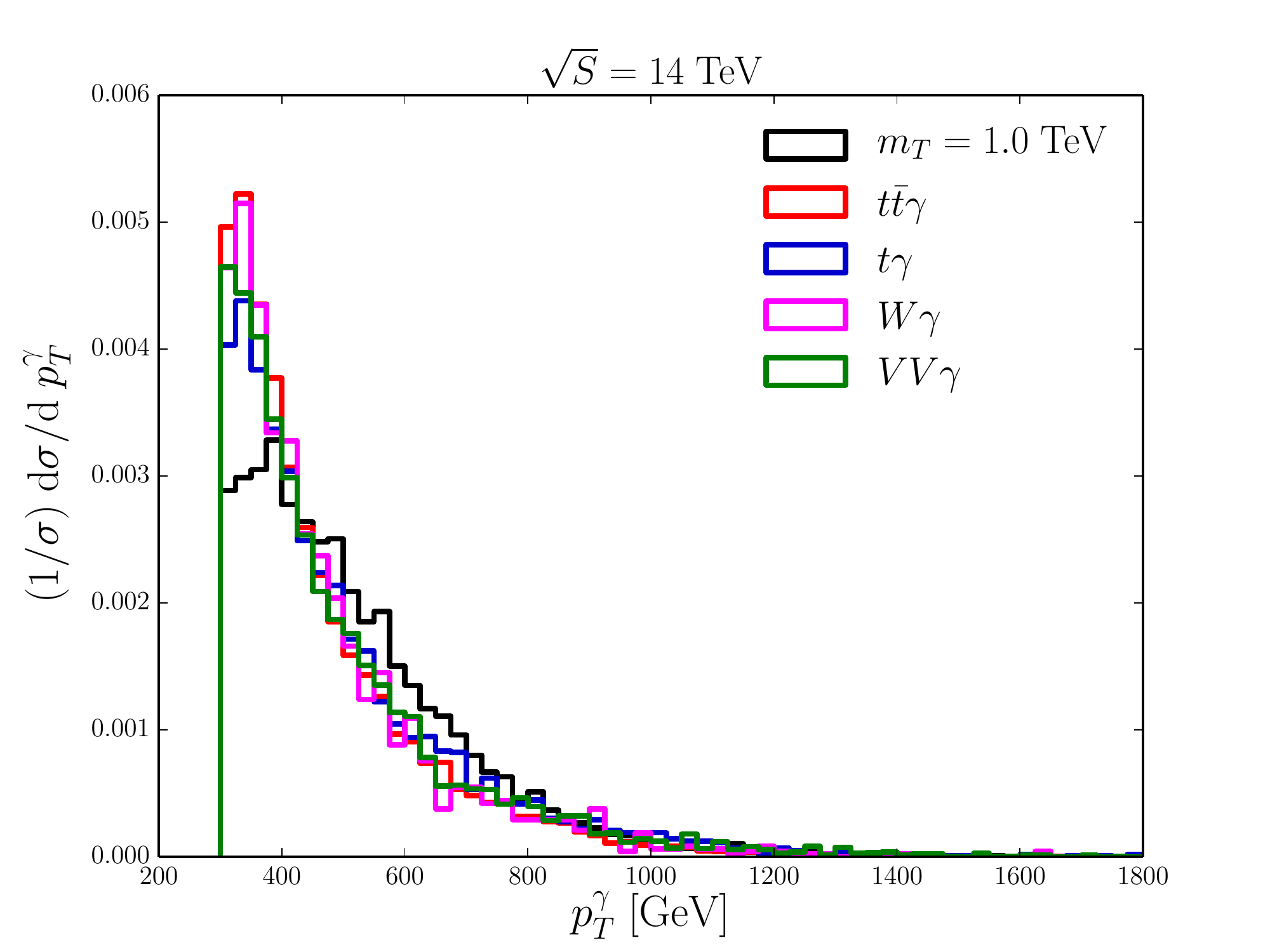}
\includegraphics[width=0.49\textwidth,clip]{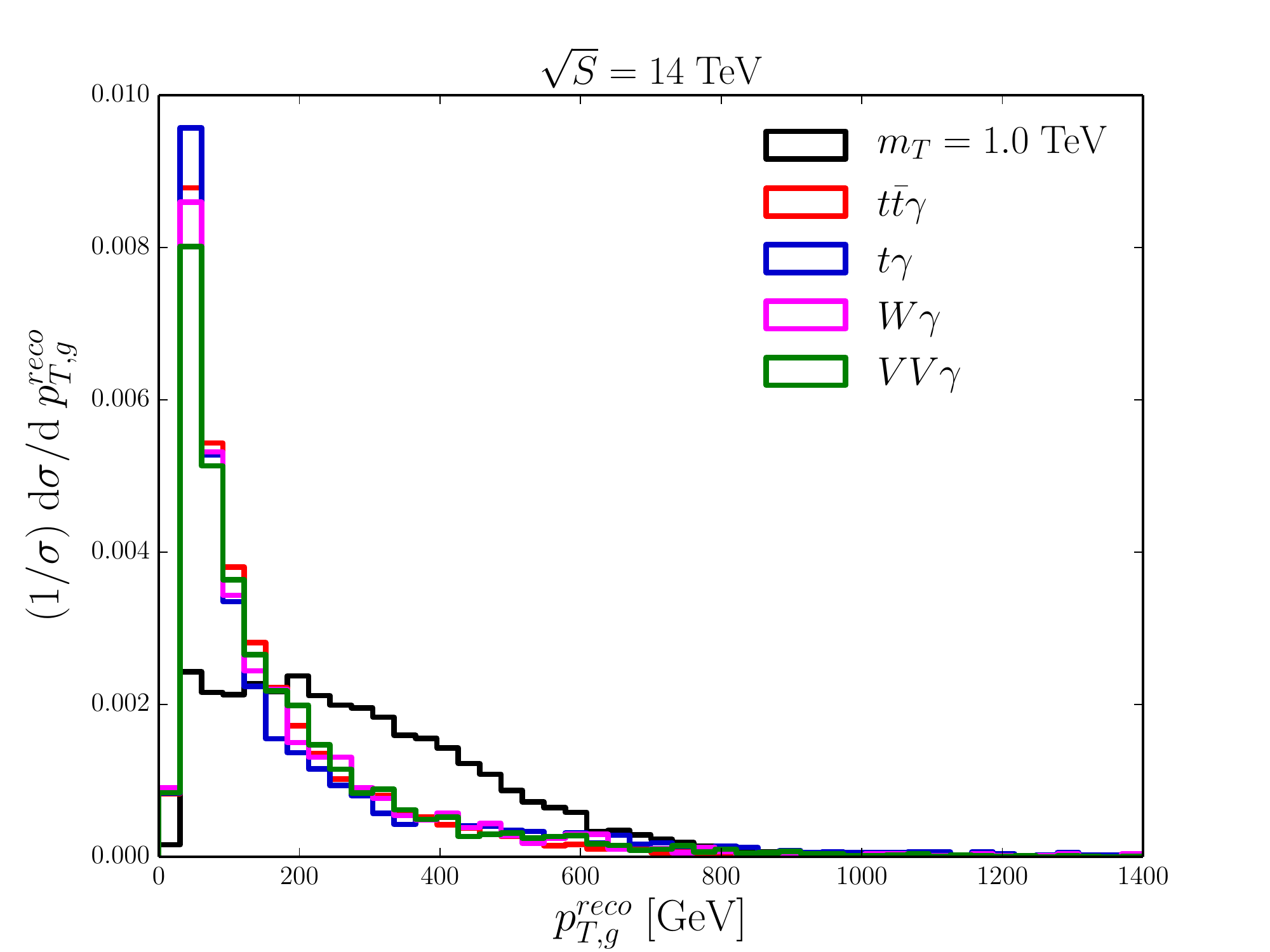}
\includegraphics[width=0.49\textwidth,clip]{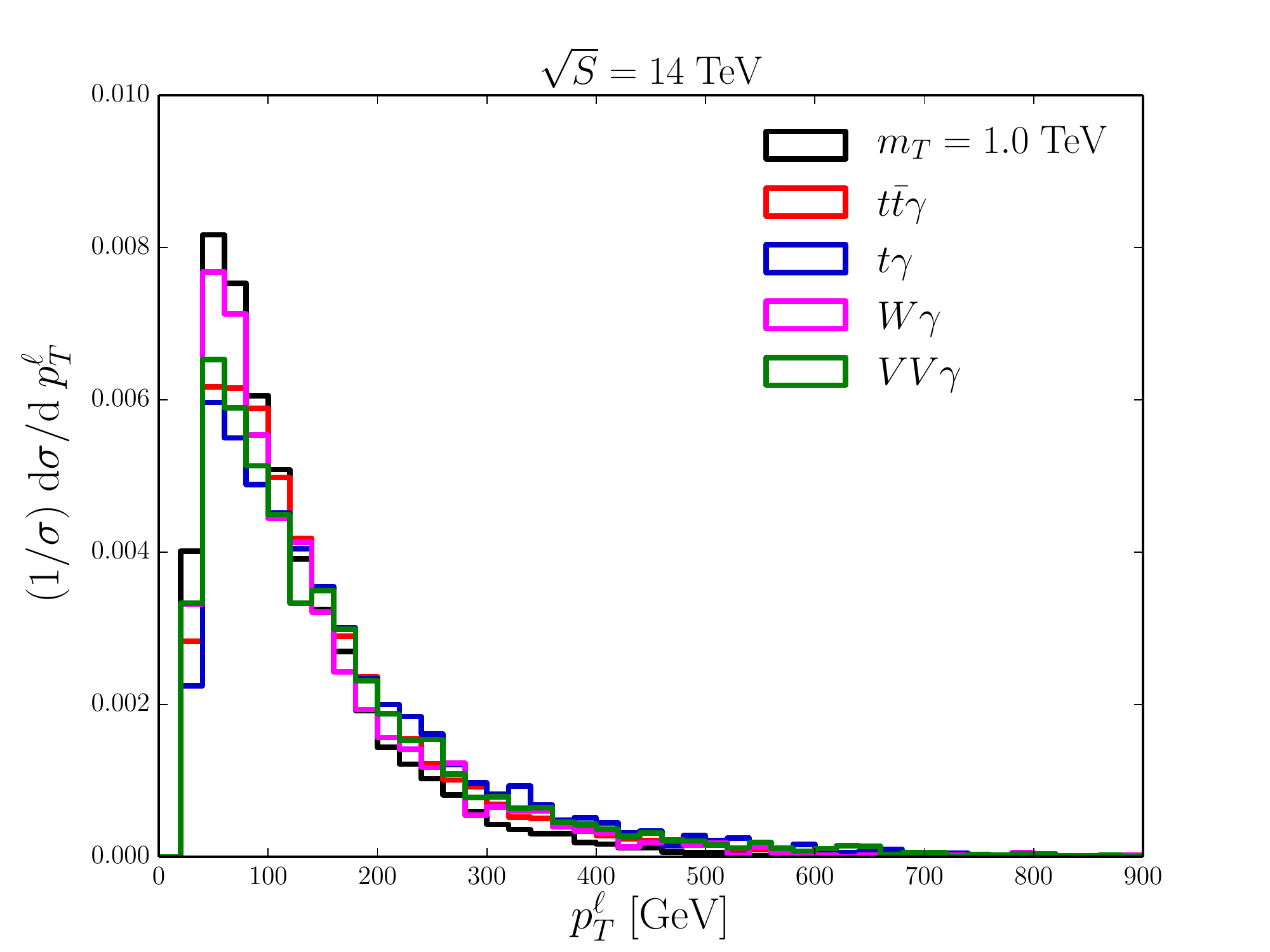}
\includegraphics[width=0.49\textwidth,clip]{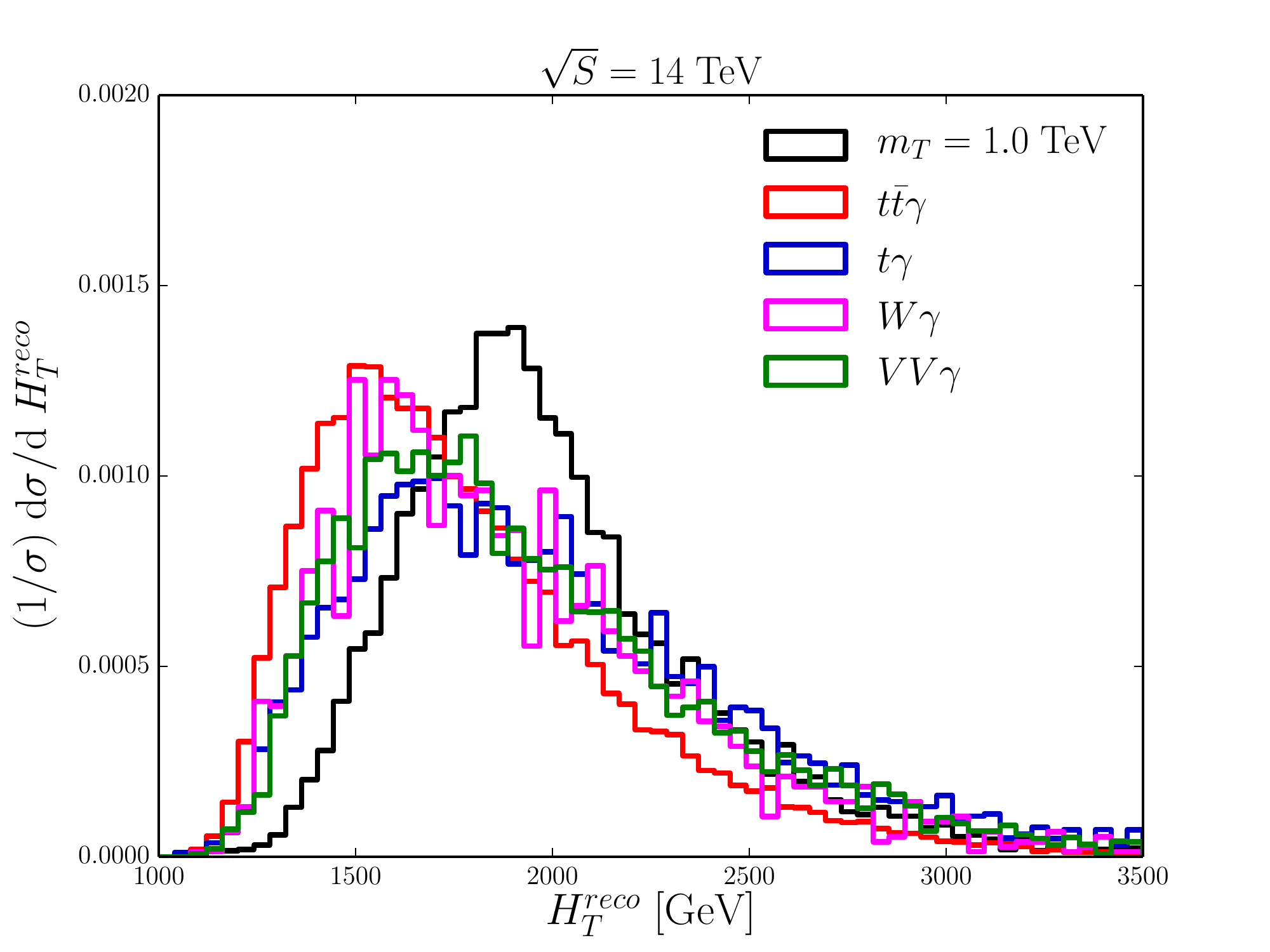}
\caption{The $p_T$ distributions of the isolated photon (top-left), the hardest slim jet not associated with $t_{had}$ or $t_{lep}$ (top-right), and the isolated lepton (bottom-left) in the $t \overline{t} g \gamma$ channel for $m_{T} = 1.0\TeV$. The scalar sum of the transverse momenta, $H_T^{reco}$ in Eq.~(\ref{eq:HTreco2}), of the reconstructed hadronic and leptonic tops, the isolated photon, and the hardest slim jet that is well-separated from the top quarks is shown in the bottom-right panel. 
\label{fig:ggmma}}
\end{figure}

Fig. \ref{fig:ggmma} shows $p_T$ distributions of the isolated photon (top-left) and the hardest reconstructed slim jet that is well-separated from the top quarks (top-right) for $m_{T} = 1.0\TeV$. The signal photons and jet have higher $p_T$ than the backgrounds and we further impose the cuts 
\begin{eqnarray}
p^{\gamma}_{T} > 300~{\rm GeV}\quad\quad{\rm and}\label{eq:PT_ga}\quad\quad
p^{reco}_{T, g} > 140~{\rm GeV}\, .
\end{eqnarray} 

Finally, we introduce a variable $H^{reco}_T$ (see the bottom-right of Fig. \ref{fig:ggmma}.) defined as the scalar sum of the transverse momenta of the reconstructed hadronic and leptonic tops, the isolated photon, and the hardest slim jet separated from the top quarks
\bea
\label{eq:HTreco2} 
	H^{reco}_T  =  p_{T,t_{had}}^{reco} + p_{T,t_{lep}}^{reco} + p^{\gamma}_{T} + p^{reco}_{T, g} \;.
\eea
The signal is much harder than the background and to obtain a higher significance we apply the cut
\begin{eqnarray}
H^{reco}_T > 1600~{\rm GeV} . \label{eq:HTCut_ga}
\end{eqnarray}

Table~\ref{tab:ttga_Cutflow_1000} is a cut-flow table showing the SM backgrounds and signal cross sections in the $t \overline{t} g \gamma$ channel for $m_{T} = 1.0$ TeV. Our result show that the outlook for the $t \overline{t} g \gamma$ channel is quite promising with a discovery significance of $\sigma_{dis} = 8.1$ at the high luminosity LHC for $m_{T} = 1.0$ TeV.

\begin{table*}[t]
\begin{center}
\setlength{\tabcolsep}{1.5mm}
\renewcommand{\arraystretch}{1.1}
\scalebox{0.84}{
\begin{tabular}{|c|ccccc|c|c|}
\hline
 $t \overline{t} g \gamma$ channel                                          & Signal [fb]               & $tt \gamma$ [fb]         & $t \gamma$ [fb]           & $W \gamma $ [fb]     &  $VV \gamma$ [fb]   & $\sigma_{dis}$ & $\sigma_{excl}$   \\ \hline \hline
Basic cuts		                                                                        & 0.13                      & 0.32                             & 1.1                               & 2.4                             & 0.10                          & 3.6                   & 3.6  \\
$N_{t_{had}}=1$                                                                      & 0.076                       & 0.22                           & 0.39                             & 0.47                           & 0.022                        & 3.9                   & 3.8 \\
$N_{t_{lep}}=1$                                                                       & 0.033                       & 0.061                         & 0.030                           & 0.029                         & $2.1 \times 10^{-3}$  & 4.9                   & 4.7  \\
$\{ p^{\gamma}_{T} , p^{reco}_{T, g} \} > \{ 300, 140\}$ GeV  & 0.021                       & 0.023                         & 0.011                          & 0.012                       & $8.8 \times 10^{-4}$  & 5.1                  & 4.7  \\
$ H_T > 1600$ GeV                                                                & 0.020                         & 0.016                         & $9.5 \times 10^{-3}$    & $9.7 \times 10^{-3}$  & $7.4 \times 10^{-4}$  & 5.2                  & 4.8  \\
\makecell{$900<m^{reco}_{T_{\gamma}}<1100\ $GeV\\ $700<m^{reco}_{T_{g}}<1100 \ $GeV}  &    0.015     & $3.1 \times 10^{-3}$  & $1.5 \times 10^{-3}$  & $1.3 \times 10^{-3}$  & $1.1 \times 10^{-4}$ & 8.1 & 6.6 \\ \hline \hline
 $b$-tag on $t_{\rm had}$                                                       & $9.6 \times 10^{-3}$ & $2.0 \times 10^{-3}$ & $7.4 \times 10^{-4}$    & $1.4 \times 10^{-4}$   & $6.1 \times 10^{-6}$  & 7.2                 & 5.7  \\  \hline \hline
 $b$-tag on $t_{\rm lep}$                                                        & $9.4 \times 10^{-3}$ & $1.8 \times 10^{-3}$ & $4.8 \times 10^{-4}$    & $2.7 \times 10^{-5}$   & $2.9 \times 10^{-6}$  & 7.6                 & 5.8  \\  \hline \hline
$b$-tag on $t_{\rm had} \; \& \;t_{\rm lep}$                             & $6.2 \times 10^{-3}$ & $1.2 \times 10^{-3}$ & $1.4 \times 10^{-4}$    & $2.1 \times 10^{-6}$   & $1.9 \times 10^{-7}$  & 6.4                 & 4.8  \\   \hline
\end{tabular}}
\end{center}
\caption{A cumulative cut-flow table showing the SM backgrounds and signal cross sections in the $t \overline{t} g \gamma$ channel for $m_{T} = 1.0$ TeV. The significances $\sigma_{dis}$ and $\sigma_{excl}$ are calculated based on the likelihood-ratio methods defined in Eq.(\ref{Eq:SigDis}) and Eq.(\ref{Eq:SigExc}) respectively for a given luminosity of $3$ ab$^{-1}$.  The summary of the background simulations can be found in Table~\ref{tab:TotalBackA}.}
\label{tab:ttga_Cutflow_1000}
\end{table*}


\subsection{Combined Analysis}\label{sec:results}

\begin{figure}[tb]
\begin{center}
\includegraphics[width=0.45\textwidth,clip]{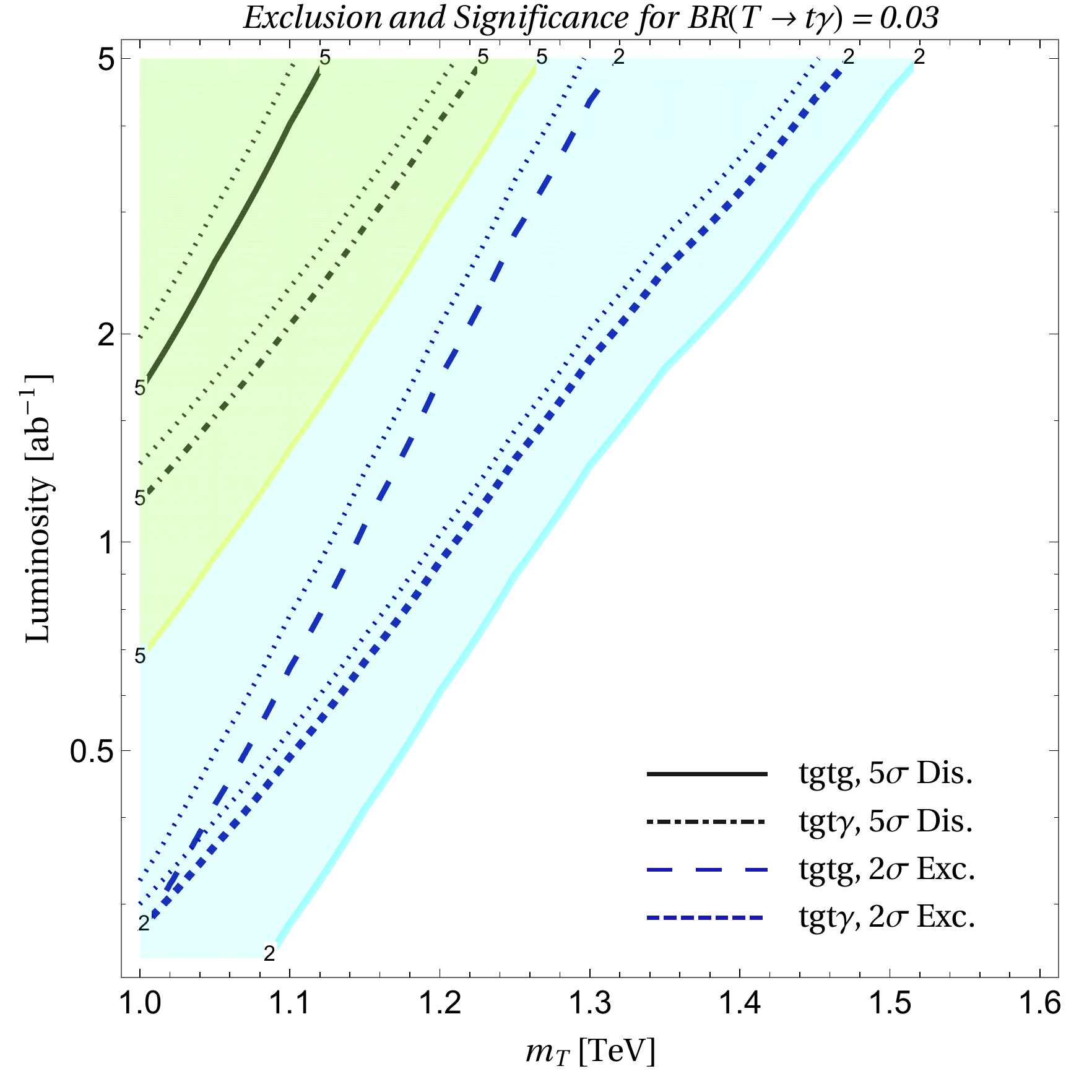} \hspace*{0.5cm}
\includegraphics[width=0.45\textwidth,clip]{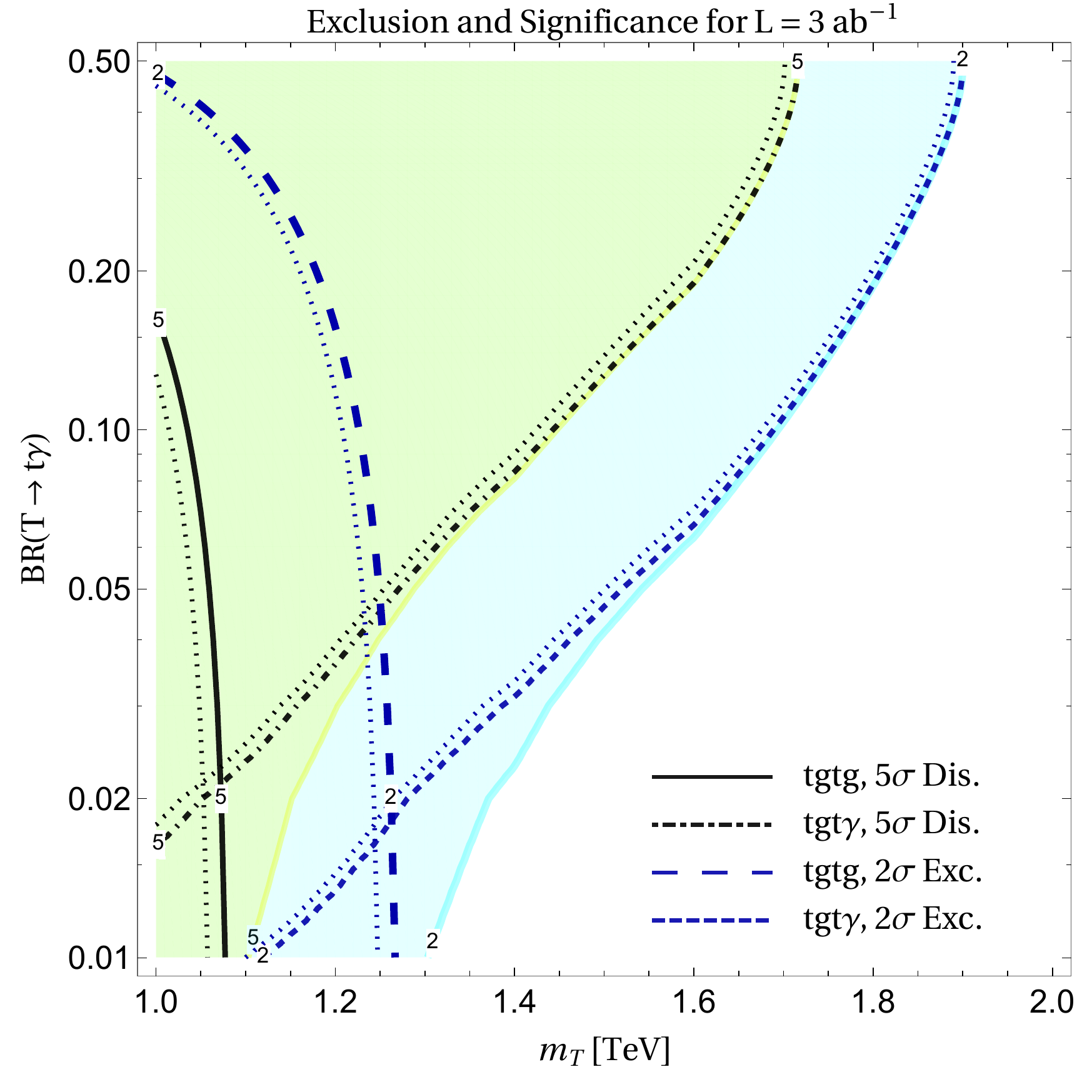}
\end{center}
\vspace*{-0.5cm}
\caption{\label{fig:spin12}
(left) The required-integrated luminosity (in ab$^{-1}$) as a function of $m_T$ for 5$\sigma$ discovery and 2$\sigma$ exclusion for ${\rm BR}(T\rightarrow t\gamma)=0.03$ and ${\rm BR}(T\rightarrow tg)=0.97$ as in Eq.~(\ref{eq:benchmark}). (right) The minimum ${\rm BR}(T \to t \gamma)$ needed for $5\sigma$ discovery and $2\sigma$ exclusion for a fixed luminosity of 3 ab$^{-1}$ as a function of $m_T$ while enforcing ${\rm BR}(T\rightarrow tg)=1-{\rm BR}(T\rightarrow t\gamma)$.  Both panels are for spin-$\frac{1}{2}$ top partner. In both plots, the 5$\sigma$ discovery result for $t\bar t g g$ ($t\bar t g \gamma$) is shown as the black-solid (black-dot-dashed) curve, while the 2$\sigma$ exclusion is shown as the blue-long-dashed (blue-short-dashed) curve. The green- and cyan-shade areas represent the combined 5$\sigma$ discovery and 2$\sigma$ exclusion, considering both $t\bar t g g$ and $t\bar t g \gamma$ channels. Dotted curves represent the corresponding results considering $20 \%$ upward fluctuation in the estimation of the background.}
\end{figure}

\begin{figure}[tb]
\begin{center}
\includegraphics[width=0.45\textwidth,clip]{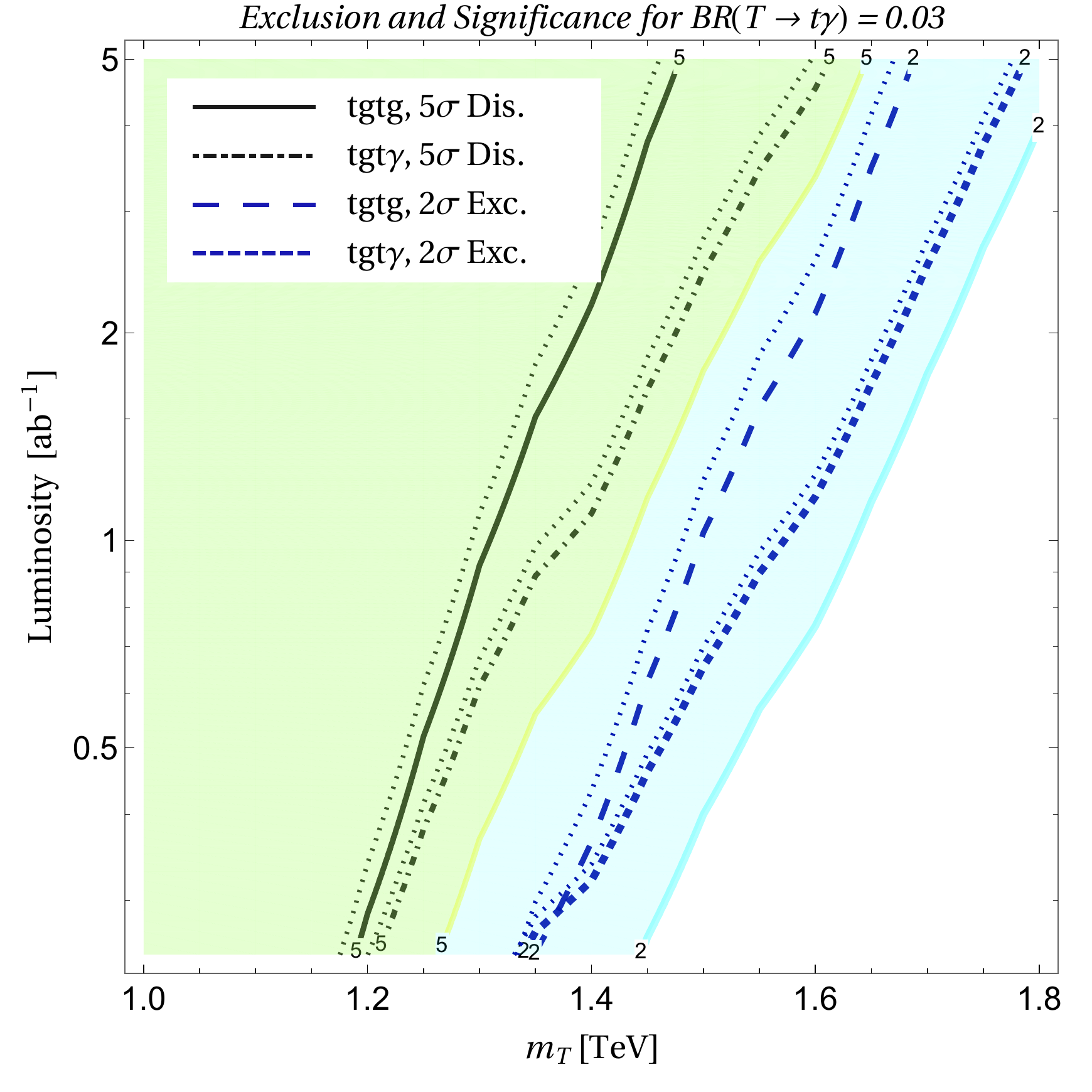} \hspace*{0.5cm}
\includegraphics[width=0.45\textwidth,clip]{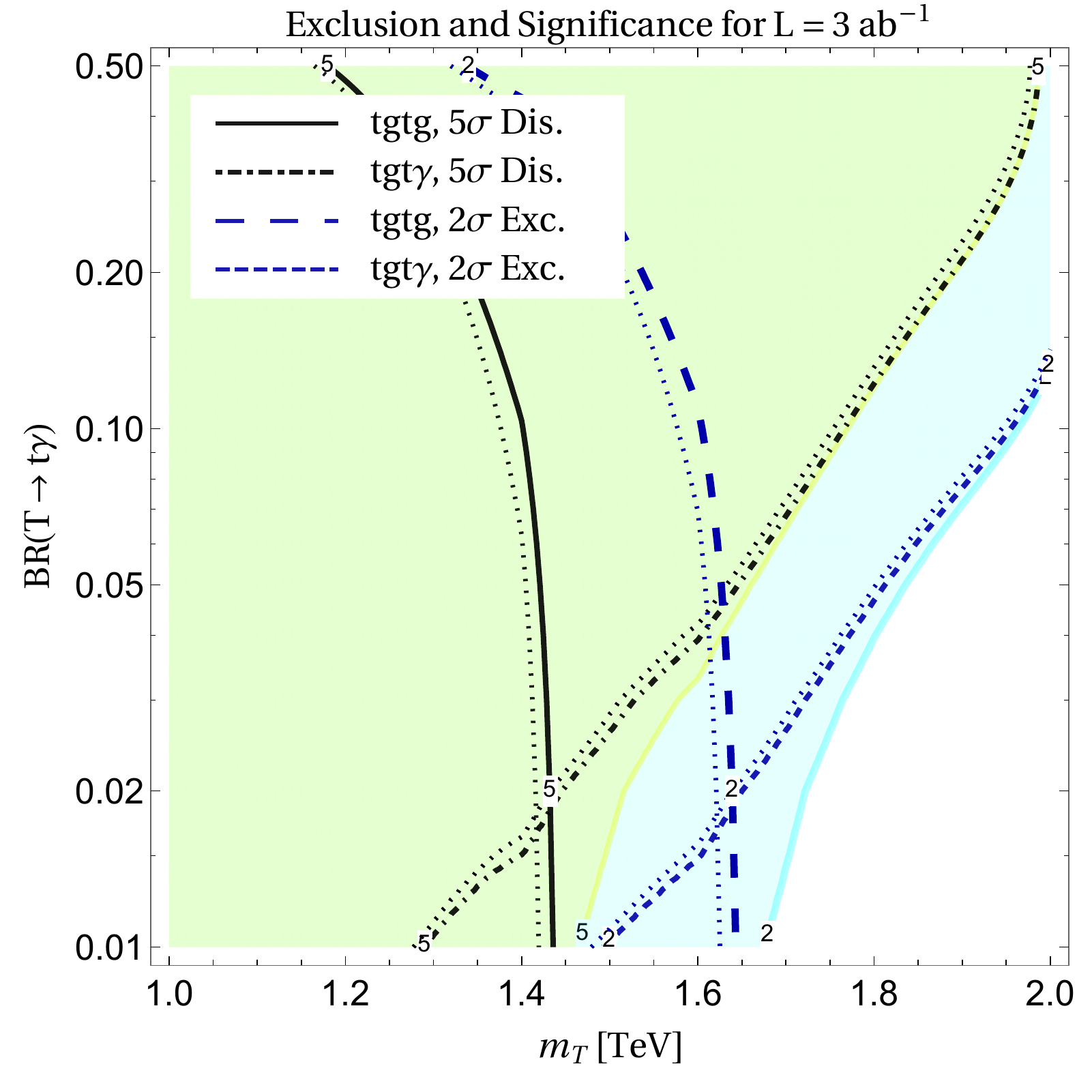}
\end{center}
\vspace*{-0.5cm}
\caption{\label{fig:spin32}The same as Fig. \ref{fig:spin12} but for spin-$\frac{3}{2}$.}
\end{figure}

In the two previous subsections, we have used a $m_T=1$ TeV spin-$\frac{1}{2}$ top partner as a benchmark model to describe our analysis and showed relevant kinematic distributions. 
We repeat similar analyses for other mass points for both spin-$\frac{1}{2}$ (Fig. \ref{fig:spin12}) and spin-$\frac{3}{2}$ (Fig. \ref{fig:spin32}) top partners.
Appendix \ref{app:CutFlow} lists optimized cuts, $\sigma_{dis}$, and $\sigma_{excl}$ for each mass point and our benchmark parameter point in Eq.~(\ref{eq:benchmark}). 

In the left panel of Fig. \ref{fig:spin12} we show the required integrated luminosity (in ab$^{-1}$) as a function of top partner mass for both a 5$\sigma$ discovery and a 2$\sigma$ exclusion of a spin-$\onehalf$ top partner.  The discovery and exclusion limits were calculated using our benchmark point of ${\rm BR}(T\rightarrow t\gamma)=0.03$ and ${\rm BR}(T\rightarrow tg)=0.97$ in Eq.~(\ref{eq:benchmark}). The right panel displays the minimum branching fraction of $T \to t \gamma$ for 5$\sigma$ discovery and 2$\sigma$ exclusion at a fixed luminosity of 3 ab$^{-1}$ while imposing ${\rm BR}(T\rightarrow t\gamma)+{\rm BR}(T\rightarrow tg)=1$. Results for spin-$\frac{3}{2}$ are shown in Fig. \ref{fig:spin32}.  We have verified that our results are consistent with the current CMS bounds excluding top-partner masses below 1.2 TeV for spin-$\frac{3}{2}$ and 930 GeV for spin-$\frac{1}{2}$ with ${\rm BR}(T\rightarrow tg)=1$.  This was accomplished by rescaling our results between 14 TeV and 13 TeV, the appropriate K-factors, etc.

In all plots, the 5$\sigma$ discovery result for $t\bar t g g$ ($t\bar t g \gamma$) is shown in black-solid (black-dot-dashed) curve, while the 2$\sigma$ exclusion is shown in blue-long-dashed (blue-short-dashed) curve.    The $t\overline{t}g\gamma$ channel is expected to have better exclusion limits for $m_T\gtrsim 1$~TeV for spin $\onehalf$ partners and $m_T\gtrsim 1.3$~TeV for spin $\threehalf$ partners.
Similarly, as shown in the right panels of Figs.~\ref{fig:spin12} and~\ref{fig:spin32}, $t\overline{t}g\gamma$ provides better signal sensitivity than $t\overline{t}gg$ for ${\rm BR}(T\rightarrow t\gamma)\gtrsim 0.02$.  It is interesting to notice that the branching fraction of $T\to t\gamma$ is expected to be a couple of percent from a naive dimensional analysis. In fact, a recent study confirmed this by explicitly computing various loop decays of $T_{\onehalf}$ in a simple model \cite{Kim:2018mks}.
There are various sources of systematic uncertainties \cite{Sirunyan:2017yta,Chatrchyan:2013oba} and we repeat the same analysis including 20\% increase (as an upward fluctuation) in the estimation of backgrounds.  The results of an upward fluctuation in backgrounds are shown in dotted curves and are essentially unchanged from the original background estimation.  
Finally, the green- and cyan-shade areas represent combined 5$\sigma$ discovery and 2$\sigma$ exclusion of both the $t\bar t g g$ and $t\bar t g \gamma$ channels.

%

\section{Summary}\label{sec:summary}

Models with top-partners are very well motivated, appearing in many BSM models.  The majority of existing analyses focus on the conventional decay modes $T\rightarrow Wb$, $T\rightarrow tZ$, and and $T\rightarrow th$, which arise due to the finite mixing between the top partner and SM top quark. As the top partner-top quark mixing angle vanishes, these decay modes are negligible and new decays become important.  In particular, loop-suppressed decays $T\rightarrow tg$ and $T\rightarrow t\gamma$ become relevant in the zero mixing angle limit~\cite{Kim:2018mks}.  

In this paper, we have investigated the discovery potential of pair-produced top-partners in the non-standard final states with gluons and photons. Using boosted techniques, we have studied the channels $T\overline{T}\rightarrow t\overline{t}+g\gamma$ and $T\overline{T}\rightarrow t\overline{t}+gg$.  The final state $t\overline{t}+g\gamma$ has not been previously studied and boosted techniques have not been previously used for the final state $t\overline{t}+gg$.  In addition to boosted techniques, we have also included relevant backgrounds with the jet-faking-photon rate. We showed that the two channels are complementary depending on the branching fraction of the top partner.   When $T \to t g$ and $T \to t \gamma$ are the two dominant decay channels, our study showed that for ${\rm BR}( T \to t \gamma)\sim\mathcal{O}(1\%)$ the $t \bar t + g\gamma$ final state has a larger significance than the $t \bar t + g g$ channel. We also showed that the combination of both channels significantly improves the signal sensitivity. With ${\rm BR}(T\to t\gamma)\sim\mathcal{O}(1\%)$, top partners can be ruled out for masses $m_T\lesssim 1.4-1.8$~TeV and discovered for masses $m_T\lesssim 1.2-1.5$~TeV for spin-$\frac{1}{2}$ and spin-$\frac{3}{2}$, respectively.   
We checked that our conclusions were stable against a 20\% upward fluctuation in the estimation of background.

Before concluding, we would like to make a brief remark on top partner searches in general.  Currently existing analyses~\cite{Sirunyan:2017pks,Sirunyan:2017ynj,Sirunyan:2018omb,Aaboud:2017zfn,Aaboud:2017qpr,Aaboud:2018saj,Aaboud:2018xuw,ATLAS-CONF-2018-032} involve final states in the entry labeled as $(1)$ in Table \ref{table:tprime} and these final states in $(1)$ assume non-negligible mixing angle between the top partner and the SM top quark, as mentioned before. If the mixing angle is small, other decay modes, such as $T\rightarrow tg$ and $T\rightarrow t\gamma$, become important and the mixed final states in $(5)$ and $(6)$ are motivated.  If the mixing angle becomes negligible, then conventional decays are closed and the only available channels would be those in $(2)$-$(4)$. The CMS collaboration \cite{Sirunyan:2017yta} started looking for spin-$\frac{3}{2}$ top partners ($T_{\threehalf}$) in the channel $(2)$ and we have advocated the channel $(3)$ in this paper. Although we argued that a simple dimensional analysis suggests a very small branching fraction to the diphoton final state as in $(4)$, this channel could have negligible backgrounds. Finally, the top-partner may interact with the SM top quark via a messenger particle $S$ and it may follow a completely different decay mode, $T \to t S$ in $(7)$-$(10)$, for example see Refs. \cite{Dolan:2016eki,Kim:2018mks,Bizot:2018tds,Dobrescu:1999gv}. Depending on the model, $S$ may decay into $gg$, $\gamma\gamma$, $g\gamma$, $WW$, $ZZ$, dark matter particles, etc.  Although Table \ref{table:tprime} illustrates possible final states in pair production, a similar classification can be easily done for single production of the top partner.  Also, it is possible to have additional exotic production and decay signatures of top partners~\cite{Kim:2018mks,Bizot:2018tds,Das:2018gcr}. We urge experimental collaborations to search for top partners in all possible final states, considering their quantum numbers and regardless of their theoretical motivations.

\begin{table}[tb]
\centering
\includegraphics[width=0.8\textwidth,clip]{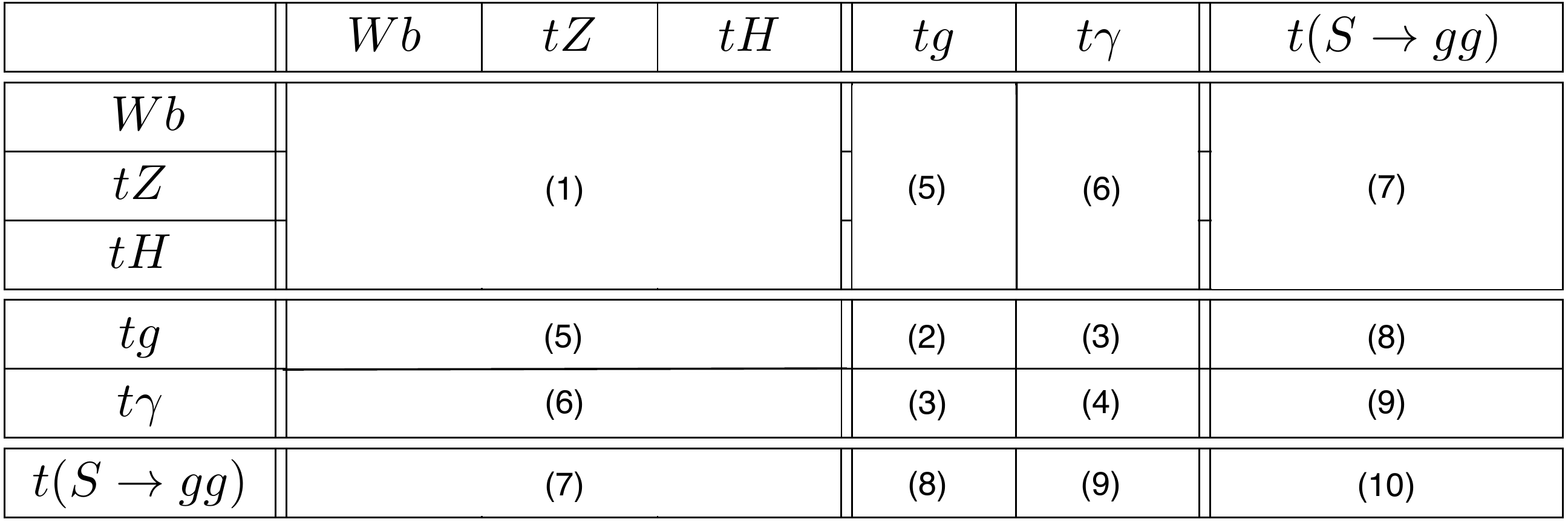}
\caption{Possible final states from the pair-produced top partner.\label{table:tprime}}
\end{table}
%


\section*{Acknowledgments}
We would like to thank Kai-Feng (Jack) Chen for helpful discussions on SM backgrounds and Sasha Pukhov for help with questions on CalcHEP. 
We also thank the University of Kansas Center for Research Computing for providing the necessary computing resources, and Riley Epperson for technical help with clusters. KK thanks the Aspen Center for Physics for hospitality during the completion of this work, supported in part by National Science Foundation grant PHY-1607611. 
This work is supported in part by United States Department of Energy (DE-SC0017988 and DE-SC0017965)  and by the University of Kansas General Research Fund
allocation 2302091. The data to reproduce the plots has been uploaded with the arXiv
submission or is available upon request.

\appendix
\section{Parameterization of Detector Resolution Effects}
\label{app:DR}

We include detector effects based on the ATLAS detector performances~\cite{ATL-PHYS-PUB-2013-004}. The energy resolution is parameterized by noise ($N$), stochastic ($S$), and constant ($C$) terms
\begin{align}
\frac{\sigma}{E} =  \sqrt{
\bigg( \frac{N}{E} \bigg)^2 +\bigg( \frac{S}{\sqrt{E}}\bigg)^2  +C^2~,
}
\end{align}
where in our analysis we use $N=5.3$, $S=0.74$ and $C=0.05$ for jets, and $N=0.3$, $S=0.1$, and $C=0.01$ for electrons; and $N=0$, $S=0.1$, and $C=0.007$ for photons~\cite{Aad:2014nim}.

The muon energy resolution is derived by the Inner Detector (ID) and Muon Spectrometer (MS) resolution functions
\begin{align}
\sigma = \frac{ \sigma_{\text{ID}}~ \sigma_{\text{MS} }  } { \sqrt{ \sigma^2_{\text{ID}} + \sigma^2_{\text{MS}   } } }~,
\end{align}
where
\bea
\label{eq:MuonSmear}
\sigma_{\text{ID}} &=& E~\sqrt{ a^2_1 + ( a_2 ~ E )^2   } \\
\sigma_{\text{MS}} &=& E~\sqrt{ \bigg( \frac{b_0}{E} \bigg)^2 + b^2_1 + (b_2~E)^2   }~\;.
\eea
We use $a_1 = 0.023035$, $a_2 = 0.000347$, $b_0 = 0.12$, $b_1 = 0.03278$ and $b_2 = 0.00014$ in our study.

\section{Summary of Cut-flow}
\label{app:CutFlow}
In Table \ref{tab:ttgg_Cutflow_mass_scan}, we summarize the cumulative cut-flow of both signal (spin-$\frac{1}{2}$ top partner) and backgrounds for various values of the top-partner mass in the $ttgg$ channel. Similar results in the $ttg\gamma$ channel are shown in Table \ref{tab:ttga_Cutflow_mass_scan}. The significance and the exclusion are calculated for a luminosity of $3000$ fb$^{-1}$.
The case with spin-$\frac{3}{2}$ top partner gives similar cut efficiencies. 

\begin{table*}[t]
\begin{center}
\def\arraystretch{1.3}
\setlength{\tabcolsep}{1mm}
\scalebox{0.77}{
\begin{tabular}{|c||c|c|c|c|c|}
\hline
\makecell{$m_T$  \\ (TeV)} & Cuts            & \makecell{$\sigma^{\text{Signal}}$ \\ (fb)}           &   \makecell{$\sigma^{\text{BG}}$ \\ (fb) }    & $\sigma_{dis}$     & $\sigma_{excl}$   \\ \hline \hline
&BC \& $t$-tagging      &~0.5999  ~   &~239.97 ~   &~2.1202 ~   &~2.1193~  \\
1.0   &~{$p^{ \{ g_1, g_2\} }_T > \{250, 150\}$ GeV   \& $ H_T > 1600$ GeV  \& $M_{t^{\text{had}}} > 145$ GeV}    ~   & 0.2932    &9.9995   &5.0545    &5.0302  \\
&{$ 750 < M_{T'} < 1100\ $ GeV}       &0.1638     &1.7214    &6.7346    &6.6333  \\   \hline  \hline
&BC \& $t$-tagging      &0.1912    &239.97     &0.6759   &0.6758  \\
1.2   &{$p^{ \{ g_1, g_2\} }_T > \{250, 150\}$ GeV  \& $ H_T > 1700$ GeV  \& $M_{t^{\text{had}}} > 145$ GeV}       &0.1207  &8.3195   &2.2860     &2.2806  \\
&{$ 950 < M_{T'} < 1300\ $ GeV}       &0.0546   &0.960    &3.022      &2.9945  \\    \hline  \hline
&BC \& $t$-tagging      &0.0644    &239.97     &0.2277     &0.2277  \\
1.4   &{$p^{ \{ g_1, g_2\} }_T > \{250, 150\}$ GeV   \& $ H_T > 1850$ GeV   \& $M_{t^{\text{had}}} > 145$ GeV}       &0.0457  &6.2833    &0.9980      &0.9968  \\
&{$ 1050 < M_{T'} < 1500\ $ GeV}       &0.0214   &0.9523    &1.1969      &1.1925  \\    \hline  \hline
&BC \& $t$-tagging       &0.0227     &239.97       &0.0804     &0.0804  \\ 
1.6   &{$p^{ \{ g_1, g_2\} }_T > \{400, 200\}$ GeV   \& $ H_T > 2100$ GeV  \& $M_{t^{\text{had}}} > 145$ GeV}       &0.0144   &2.3249      &0.5168     &0.5162  \\
&{$ 1100 < M_{T'} < 1800\ $ GeV}       &9.26e-3    &0.7409    &0.5883     &0.5871  \\    \hline  \hline
&BC \& $t$-tagging      &8.093e-3     &239.97   &0.0286      &0.0286  \\
1.8   &{$p^{ \{ g_1, g_2\} }_T > \{500, 200\}$ GeV  \& $ H_T > 2350$ GeV  \& $M_{t^{\text{had}}} > 145$ GeV}       &5.12e-3        &1.3104     &0.2449      &0.2448  \\
&{$ 1150 < M_{T'} < 2100\ $ GeV}       &3.59e-3      &0.5326    &0.2683   &0.2680  \\    \hline  \hline
&BC \& $t$-tagging      &2.94e-3   &239.97     &0.010e   &0.0104  \\
2.0   &{$p^{ \{ g_1, g_2\} }_T > \{500, 200\}$ GeV  \& $ H_T > 2500$ GeV  \& $M_{t^{\text{had}}} > 145$ GeV}      &1.95e-3  &0.9403    &0.1104    &0.1103  \\
&{$ 1150 < M_{T'} < 2500\ $ GeV}       &1.53e-3   &0.4521   &0.1252   &0.1251  \\    \hline
\end{tabular}}
\end{center}
\caption{Cumulative cut-flow in the $ttgg$ channel for both signal (spin-$\frac{1}{2}$ top partner) and backgrounds. 
The significance and the exclusion are calculated for a luminosity of $3000$ fb$^{-1}$. The case with spin-$\frac{3}{2}$ top partner is similar.}
\label{tab:ttgg_Cutflow_mass_scan}
\end{table*}


\begin{table*}[t]
\centering
\begin{center}
\def\arraystretch{1.3}
\setlength{\tabcolsep}{1mm}
\scalebox{0.77}{
\begin{tabular}{|c||c|c|c|c|c|}
\hline
\makecell{$m_T$  \\ (TeV)} & Cuts            & \makecell{$\sigma^{\text{Signal}}$ \\ (fb)}           &   \makecell{$\sigma^{\text{BG}}$ \\ (fb) }    & $\sigma_{dis}$     & $\sigma_{excl}$   \\ \hline \hline
 &BC \& $t$-tagging   &0.0295   &0.1364      &4.2294     &4.0937 \\ 
1.0 & ~{$p^{ \{ \gamma, g\} }_T > \{300, 140\}$ GeV   \& $ H_T > 1600$ GeV   \& $M_{t^{\text{had}}} > 145$ GeV} ~ &~0.0196~    &~0.0365~      &~7.5814~    &~6.8712~ \\ 
& {$ 900 < M^{\gamma}_{T'} < 1100\ $ GeV $ 700 < M^{g}_{T'} < 1100\ $ GeV}      &0.0147    & 6.02e-3    &8.0807    &6.5927  \\ \hline \hline
&BC \& $t$-tagging     &0.0115    &0.1364           &1.6779     &1.6555   \\ 
1.2 & {$p^{ \{ \gamma, g\} }_T > \{300, 150\}$ GeV   \& $ H_T > 2000$ GeV  \& $M_{t^{\text{had}}} > 145$ GeV}       &6.67e-3     &0.0189         &4.1985      &3.8795   \\ 
& {$ 1100 < M^{\gamma}_{T'} < 1300\ $ GeV  $ 850 < M^{g}_{T'} < 1300\ $ GeV}     &4.72e-3    &2.34e-3         &4.2913     &3.5768     \\   \hline  \hline
&BC \& $t$-tagging        &4.204e-3    &0.1364    &0.6204     &0.6173  \\ 
1.4 & {$p^{ \{ \gamma, g\} }_T > \{300, 150\}$ GeV   \& $ H_T > 2200$ GeV  \& $M_{t^{\text{had}}} > 145$ GeV}          &2.70e-3     &0.0134     &1.2376    &1.2004    \\ 
&{$ 1250 < M^{\gamma}_{T'} < 1500\ $ GeV  $ 1000 < M^{g}_{T'} < 1550\ $ GeV}               &1.9185e-3    &1.7438e-3       &2.1896    &1.9358    \\ \hline   \hline
&BC \& $t$-tagging     &1.515e-3      &0.1364     &0.2243     &0.2239   \\ 
1.6 & {$p^{ \{ \gamma, g\} }_T > \{300, 200\}$ GeV   \& $ H_T > 2300$ GeV   \& $M_{t^{\text{had}}} > 145$ GeV}     &1.0428e-3      &0.0103      &0.5534     &0.5446   \\ 
&{$ 1400 < M^{\gamma}_{T'} < 1700\ $ GeV $ 1000 < M^{g}_{T'} < 1700\ $ GeV}        &7.7161e-4     &1.646e-3    &0.9730     &0.9127  \\ \hline \hline
&BC \& $t$-tagging     &5.65e-4      &0.1364     &0.0837      &0.0837  \\ 
1.8 & \makecell{$p^{ \{ \gamma, g\} }_T > \{300, 200\}$ GeV     \& $ H_T > 2600$ GeV    \&$M_{t^{\text{had}}} > 145$ GeV}    &3.83e-4    &5.47e-3    &0.4113   &0.4046    \\ 
&{$ 1700 < M^{\gamma}_{T'} < 1900\ $ GeV  $ 1100 < M^{g}_{T'} < 2000\ $ GeV}       &2.36e-4    &4.15e-4     &0.5865     &0.5441     \\  \hline   \hline
&BC \& $t$-tagging      &2.07e-4     &0.1364      &0.0308      &0.0308  \\ 
2.0 & $p^{ \{ \gamma, g\} }_T > \{300, 150\}$ GeV   \& $ H_T > 2700$ GeV  \& $M_{t^{\text{had}}} > 145$ GeV         &1.54e-4     &4.81e-3     &0.1210    &0.1203   \\ 
& $ 1800 < M^{\gamma}_{T'} < 2100\ $ GeV  \& $ 1300 < M^{g}_{T'} < 2100\ $ GeV               &9.51e-5     &2.90e-4    &0.2911     &0.2777    \\ \hline
\end{tabular}}
\end{center}
\caption{The same as Table \ref{tab:ttgg_Cutflow_mass_scan} but for the $ttg \gamma$ channel.}
\label{tab:ttga_Cutflow_mass_scan}
\end{table*}

\bibliographystyle{JHEP}
\bibliography{draft}

\end{document}